\documentclass{aa}

\usepackage{txfonts}
\usepackage{graphicx}
\usepackage{booktabs}
\usepackage{subcaption}
\usepackage{longtable}   
\usepackage{lscape}
\usepackage{xcolor}
\usepackage{enumitem}
\usepackage[nice]{nicefrac}
\usepackage{svg}
\usepackage{lastpage}
\usepackage{hyperref}   
\begin{document}

  \title{Predisposition of galaxy clusters to producing exotic hyperbolic umbilic lensing configurations}

  \author{
      Quentin Basto\inst{1} \and
      Johan Richard\inst{1} \and
      David J. Lagattuta\inst{2,3,4} \and
      Ashish K. Meena\inst{5} \and
      Harald Ebeling\inst{6}
  }

  \institute{
      Univ Lyon, Univ Lyon1, Ens de Lyon, CNRS, Centre de Recherche Astrophysique de Lyon UMR5574, F-69230, Saint-Genis-Laval, France \and
      Centre for Astrophysics Research, Department of Physics, Astronomy and Mathematics, University of Hertfordshire, Hatfield AL10 9AB, UK \and
      Centre for Extragalactic Astronomy, Durham University, South Road, Durham DH1 3LE, UK \and
      Institute for Computational Cosmology, Durham University, South Road, Durham DH1 3LE, UK \and
      Department of Physics, Indian Institute of Science, Bengaluru 560012, India \and
      Institute for Astronomy, University of Hawaii, 640 N Aohoku Pl, Hilo, HI 96720, USA
  }

  \date{Received xxx; accepted xxx}

   \abstract
  {Strong gravitational lensing is a powerful tool for investigating the universe's large-scale structure and understanding the properties of dark matter and dark energy. The magnification and distortion of distant background sources by cluster lenses have enabled detailed studies of both lens and source populations, making these systems promising probes for precision cosmology.}
  {While classical strong-lenses are well understood, much remains to be explored for hyperbolic-umbilic (HU) exotic lenses, which produce unique telescopic effects and uncommon images with potentially very high magnifications. Identifying and quantifying these objects, along with characterising their geometric configurations, could have broad implications for studies of galaxy clusters and lensed galaxy populations.}
  {Using parametric cluster mass models, we mapped regions in the source plane where HU exotic images can form and integrate these areas over redshift to define an exotic comoving volume \(V_{z<10}\). We validated this approach on confirmed exotic systems (RXJ0437.1+0043 and Abell 1703), then applied it to a sample of 74 cluster models.}
  {We show HU-region contours for the most promising clusters, assess both systematic and stochastic uncertainties on exotic area and volume estimates, and confirm that our error remains sufficiently small to support robust conclusions. Next, we explore correlations between six cluster parameters and \(V_{z<10}\), finding that pairs of parameters, especially ellipticity with Einstein radius or cuspiness, best distinguish high-\(V_{z<10}\) systems. Finally, we estimate that each cluster contributes \(\approx0.125\) galaxies to its exotic volume on average (as a conservative lower bound), meaning that observing 19 clusters yields a 90\% chance of detecting at least one HU system in a random sample.}


   \maketitle

\section{Introduction}

    Strong gravitational lensing has become a fundamental tool in astrophysics, offering profound insights into the mass distribution of galaxy clusters and acting like natural telescopes for observing distant and faint sources. Near critical lines, gravitational lenses create significant magnification \citep{natarajan1997}, forming classical arcs and enabling the study of both background sources and the internal structure of the lenses themselves \citep{Lagattuta2023}. Galaxy clusters, with their complex, dark matter-dominated mass distributions, serve as ideal laboratories for these studies.

    While classical strong-lenses are already well understood and documented, a rare and intriguing subclass, known as exotic lenses, presents unique observational and theoretical opportunities. Among these, hyperbolic umbilic (HU) lenses stand out due to their exceptional characteristics \citep{Xivry2009, meena2020}. HU lenses produce unique configurations of multiple images: typically four primary images offset from the cluster centre, plus a counter-image, with predominantly isotropic distortions and magnification factors often exceeding 100 \citep{limousin2008}. These properties allow detailed observations of distant regions of the Universe, unveiling even substructures in background sources and provide ideal constraints for lens modelling and dark matter studies \citep{Lagattuta2023}. 

    Despite their promise, HU lenses remain rare, as they require precise geometric alignments among the lens, the source, and the observer, as well as finely tuned redshift combinations \citep{schneider1992, petters2012singularity}. Estimates of their all-sky abundance vary: early works predict a single observable HU system \citep{Xivry2009}, while more recent forecasts suggest that a handful may be uncovered by current and upcoming surveys \citep{meena2021}. The factors that make a given cluster more likely to produce such rare configurations remain unclear, motivating a systematic exploration of the physical drivers behind these exotic image formations. Understanding these conditions could provide deeper insights into both the formation of gravitational lensing phenomena and the structural characteristics of galaxy clusters.

    The recent advent of \emph{JWST} deep-field observations \citep{zitrin2026strong} and wide-filed Euclid programs \citep{bergamini2025euclid} marks a turning point in strong-lensing studies. With unprecedented depth, spatial resolution, and multi-band coverage, these datasets will expand the strong-lens census by orders of magnitude and reveal a large number of faint systems, creating new opportunities to uncover new HU exotic image configurations. Since in-depth follow-up of every cluster to search for faint HU image formations is impractical, it is essential to develop automated, high-throughput pipelines able to flag and prioritise the most promising clusters to guide further detailed investigations.

    In this paper, we present a novel method to locate the regions within galaxy clusters where HU lensing is most likely to occur using parametric mass models and ray-tracing simulations. We apply our pipeline to 74 strong-lensing clusters showing a variety of lensing strength and morphology to identify key characteristics of the most promising ones for hosting exotic lenses configurations. Ultimately, this work aims to bridge the gap between theoretical predictions and observational discoveries of HU lenses, contributing to the broader field of gravitational lensing studies. 

    The rest of the paper is organised as follows. Section~\ref{sec:identification} describes the cluster sample and our HU detection algorithm. Results on the cluster sample are detailed in Section~\ref{sec:application}. Section~\ref{sec:discussions} discusses the method's robustness, the properties of the most promising clusters and the expected HU abundance in observations. We summarise our main conclusions in Section~\ref{sec:conclusion}.

    Throughout this paper we make use of magnitudes in the AB photometric system when relevant \citep{oke1983secondary}, and we adopt a standard \(\Lambda\)CDM cosmology with \(\Omega_{m}=0.3\), \(\Omega_{\Lambda}=0.7\), and \(H_{0}=70\,\mathrm{km\,s^{-1}\,Mpc^{-1}}\) for all lensing calculations.

\section{Identification of hyperbolic-umbilics and HU exotic images}
\label{sec:identification}
    In this section, we detail the detection of HU locations in both the lens and source planes and explain how we predict the regions where HU exotic image formations occur. Our input data are parametric descriptions of cluster mass distributions, which we use to generate gravitational-lensing maps. Specifically, we employ Lenstool\footnote{https://git-cral.univ-lyon1.fr/lenstool/lenstool} \citep{jullo2007} to produce convergence maps (\(\kappa\)), quantifying isotropic deformations within the lens, and shear maps (\(\gamma_1, \gamma_2\)), capturing anisotropic distortions-i.e. image stretching or compression along the \(x\)-axis (\(\gamma_1\)) and \(y\)-axis (\(\gamma_2\)) in the lens plane. The output maps cover \(200 \times 200~\mathrm{arcsec}^2\) (to include the main strong-lensing region for all clusters) sampled at 0.066 arcsec / pixel.

\subsection{Hyperbolic-umbilic configurations and detection algorithm}\label{sec:HU_detection}

\begin{figure}[!h]
        \centerline{\includegraphics[width=\columnwidth]{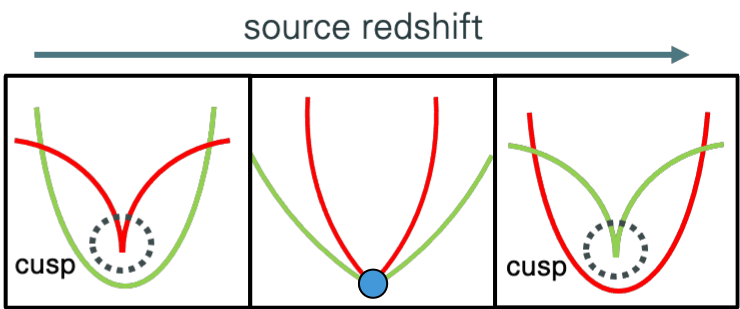}}
        \caption{Diagram illustrating the evolution of tangential (red) and radial (green) caustic lines with source redshift and the location of a hyperbolic umbilic (HU) point at the cusp exchange (blue). The dashed circles highlight the cusp-exchange point in the source plane. HU configurations are characterised by the formation of a compact quadruplet of highly magnified images at the cusp exchange, accompanied by an additional counter-image, which motivates the requirement of at least five lensed images. Adapted from \citet{petters2012singularity}.}
        \label{schema_hu}
    \end{figure}

    Hyperbolic umbilic (HU) configurations arise when the image of a radial cusp and the image of a tangential cusp meet in the lens plane, a situation that occurs only at a specific source redshift \citep{Xivry2009, meena2020}. A cusp here refers to a singular point of the caustic curve in the source plane where two folds join, corresponding to the merging of three highly magnified images \citep{schneider1992}. Consequently, the HU point appears when the transition from a radial to a tangential cusp happens, at a given source redshift (Figure~\ref{schema_hu}).

\begin{figure*}
    \sidecaption
    \includegraphics[width=12cm]{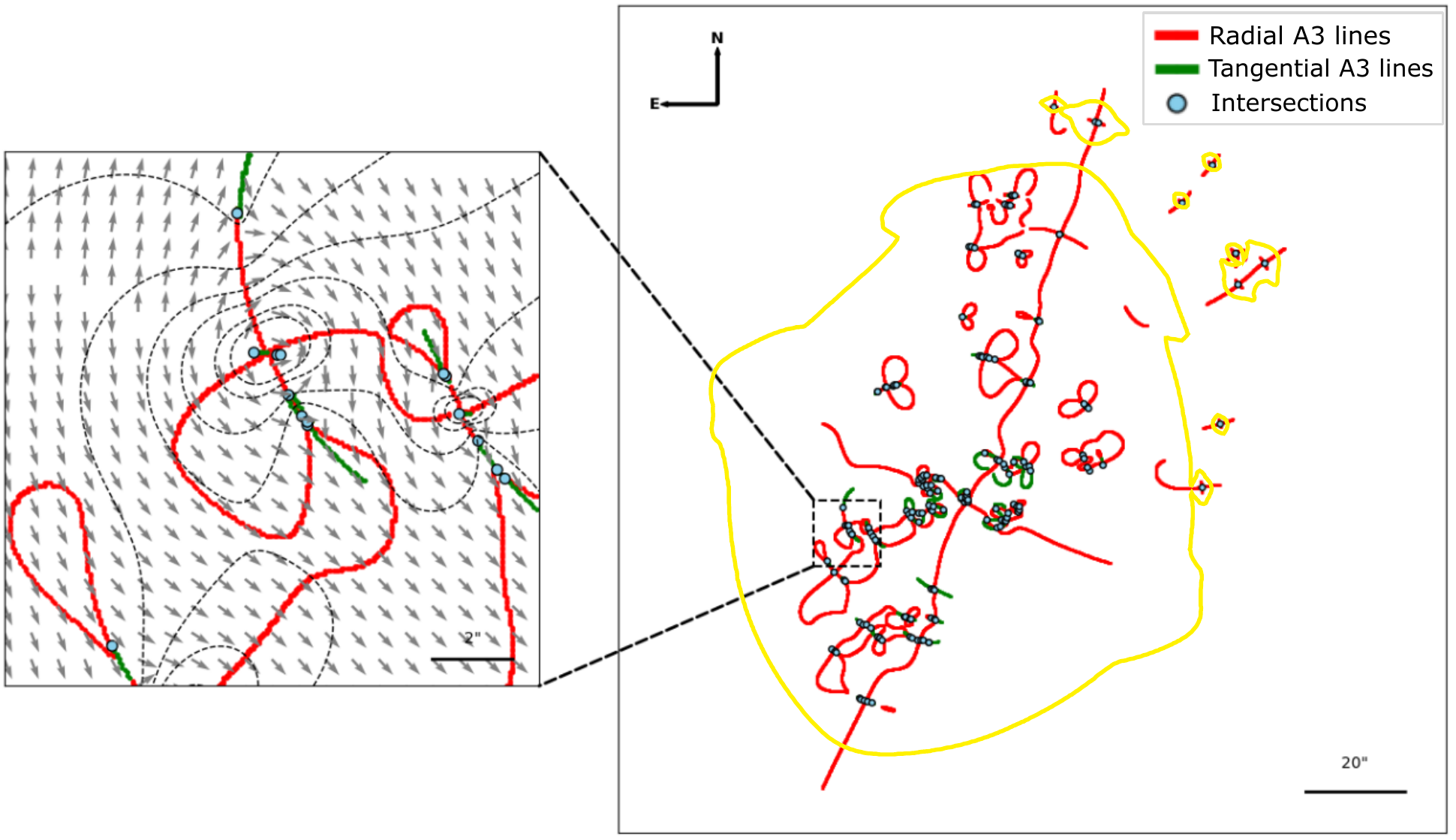}
    \caption{A3 line construction for A1703. Tangential and radial A3 lines are shown in red and green, respectively, with HU points in light blue. The close-up highlights shear vectors and $\kappa + \gamma$ isocontours (dashed black), where radial A3 lines occur where the eigenvalue isocontour and the shear vector are aligned. The yellow contour outlines the multiple-image region of the cluster, computed at an effective source redshift of $z=100$ to approximate the $D_{ls}/D_s \simeq 1$ limit.} 
    \label{a3_schema_a1703} 
    \end{figure*}

    To detect HU points in practice, we constructed the so-called A3 lines \citep{meena2021}. These curves follow the motion of cusp images across the lens plane as the source redshift, and hence the geometric factor \(D_{LS}/D_{S}\), changes. Radial cusps generate radial A3 lines, and tangential cusps generate tangential A3 lines. HU points appear precisely where the two types of A3 lines intersect, marking the locus of the cusp exchange in the image plane (Figure~\ref{a3_schema_a1703}).

    The deformation of a lensed image can be described by the Jacobian matrix of the lens equation, 
    \[
      \mathbf{A} = \delta_{ij} - \psi_{ij},
    \]
    where the coefficients 
    \[
      \psi_{ij} = \left( \frac{D_{ls}}{D_s} \right)
      \begin{pmatrix} 
      \kappa + \gamma_1 & \gamma_2 \\[1mm] 
      \gamma_2 & \kappa - \gamma_1 
      \end{pmatrix}
    \]
    define the so-called deformation tensor of the lens. The eigenvalues of the deformation tensor are 
    \[
    \alpha = \left(\frac{D_{ls}}{D_s}\right)(\kappa + \gamma), 
    \quad 
    \beta = \left(\frac{D_{ls}}{D_s}\right)(\kappa - \gamma),
    \]
    with the shear amplitude 
    \(\gamma = \sqrt{\gamma_1^2 + \gamma_2^2}.\) The lensing magnification is then given by the inverse determinant of the Jacobian matrix,
    \[
    \mu = \frac{1}{\det \mathbf{A}} = \frac{1}{(1 - \alpha)(1 - \beta)}.
    \]

    In practice, the A3 lines pass through points where the gradient of one or both eigenvalues \(\lambda\) is orthogonal to its associated eigenvector \(\mathbf{n}_\lambda\). Consequently, the condition is written as

    \begin{equation}
    \mathbf{n}_\lambda \cdot \nabla \lambda = 0.
    \label{eq:orthogonal}
    \end{equation}

    Because the convergence \(\kappa\) only scales the deformation isotropically, the eigenvectors of the deformation tensor \(\psi_{ij}\) are aligned with those of the shear tensor, which governs the anisotropic part of the lensing. The shear components can be expressed in polar form
    \(\gamma_1 = \gamma \cos2\phi\) and \(\gamma_2 = \gamma\sin2\phi\), where \(\phi\) is the shear angle. Diagonalising the shear matrix shows that the eigenvector associated with the eigenvalue \(\gamma\) is oriented as the angle \(\phi\), defining the shear direction.

    As a result, the algorithm to construct the A3 lines from the convergence and shear maps (\(\kappa\), \(\gamma_1\), \(\gamma_2\)) followed these steps:

    \begin{enumerate}
        
        \item Eigenvalue maps: we computed the $\alpha$ and $\beta$ maps. Then, we calculated the vertical
        ($\nabla_y \lambda$) and horizontal ($\nabla_x \lambda$) components of
        $\nabla \lambda$ (for both $\alpha$ and $\beta$) using a Sobel filter.

        \item Shear orientation: we computed the oriented shear angle $\phi$, defined between $[-\frac{\pi}{2};\frac{\pi}{2}]$, as \(\phi = \frac{1}{2} \text{atan2}(\gamma_2, \gamma_1)\).

        \item A3 condition: we identified points where the dot product between the shear direction and the eigenvalue gradient vanishes: $\mathbf{\nabla} \lambda \cdot \boldsymbol{\phi} = 0$. In practice, sign changes between adjacent pixel in the dot product map were used to locate the A3 lines.

        \item Thresholding: The A3 lines correspond to cusp images for which the eigenvalues satisfy
        \[
        \kappa + \gamma = \frac{D_s}{D_{ls}} \quad \text{(tangential)} \qquad \text{or} \qquad
        \kappa - \gamma = \frac{D_s}{D_{ls}} \quad \text{(radial)}.
        \]
        Since by definition $D_s > D_{ls}$, the right-hand side of these relations is strictly larger than unity. As a consequence, pixels for which $\kappa + \gamma \leq 1$ or $\kappa - \gamma \leq 1$ cannot satisfy the critical condition and were therefore discarded. This filtering step removed regions that cannot physically host cusp images and prevents spurious detections associated with subcritical or numerically unstable pixels.

        \item HU points: finally, HU points were identified as the intersections of radial and tangential A3 lines. As caustic cusps are associated with diverging magnification ($\mu \to \infty$), one of the following conditions must be met:
        \begin{equation}
        1 - \alpha = 0 \quad \text{or} \quad 1 - \beta = 0 .
        \label{eq:condition_hu_1}
        \end{equation}

        Theoretically, a cusp corresponds to a local degeneracy of the Jacobian where the two eigenvalues coincide, implying a vanishing shear ($\gamma = 0$), and therefore
        \begin{equation}
        \alpha = \beta = \frac{D_{ls}}{D_s}\kappa = 1
        \quad \Rightarrow \quad \frac{D_{ls}}{D_s} = \frac{1}{\kappa},
        \label{eq:condition_hu_2}
        \end{equation}
        which yields a unique source redshift in the continuous mapping. In practice, lensing quantities were evaluated on a discrete grid. At the adopted resolution of 0.066 arcsec per pixel, the shear does not vanish exactly at HU locations due to the numerical evaluation of second derivatives of the lensing potential. In the vicinity of HU regions, we measured typical pixel-to-pixel variations of the shear of order $\langle |\Delta\gamma| \rangle_{\rm HU} \sim 10^{-3}$--$10^{-2}$, with rms fluctuations reaching a few $10^{-2}$ in the most structured cases. These values provided a direct empirical estimate of the effective local accuracy with which \textsc{Lenstool} computes $\gamma$ at this resolution.

        As a consequence, two estimates of $\frac{D_{ls}}{D_s}$ were obtained from the two eigenvalues of the Jacobian, which did not coincide exactly at the pixel level. HU points were retained only if both estimates deviated by at most $20\%$ from the theoretical value $\frac{1}{\kappa}$. This tolerance was selected after several tests as a compromise that robustly recovers HU configurations previously reported in the literature, while avoiding excessive fragmentation or contamination by spurious candidates. Importantly, this choice was also consistent with the effective numerical precision of the lensing maps: the adopted $20\%$ tolerance is conservative with respect to the measured local fluctuations of the shear and reflects the finite accuracy with which the theoretical HU condition can be enforced on discretised maps. The final estimate of $\frac{D_{ls}}{D_s}$ for each HU point was then taken as the average of the two values, and the corresponding source redshift was assigned accordingly.

    \end{enumerate}

    To assess the accuracy of A3 line tracing and HU point identification, we applied our algorithm to complex cluster models, including Abell 1703, hereafter A1703 \citep{limousin2008}, whose results are illustrated on Figure \ref{a3_schema_a1703}, with radial and tangential A3 lines shown in red and green, respectively, with their intersections marking HU points in blue. The zoomed-in window offers a detailed view of a subregion of the map, illustrating the construction of radial A3 lines with normalised shear vectors and isocontours of the eigenvalue \(\alpha\).

    Once the HU points are known, our goal was to assess whether a source located at a given position and at a specific redshift is likely to produce an HU exotic lensing configuration. This involves estimating:
    \begin{itemize}
        \item[$\bullet$] The surface area around background sources where HU exotic images are likely to form.
        \item[$\bullet$] The redshift interval within which such configurations can occur.
    \end{itemize}
    These aspects are further explored in the following sections, with a focus on the construction of these exotic regions and their implications for observational strategies.

    \begin{figure*}[!h]
      \centering
      \includegraphics[width=0.32\textwidth]{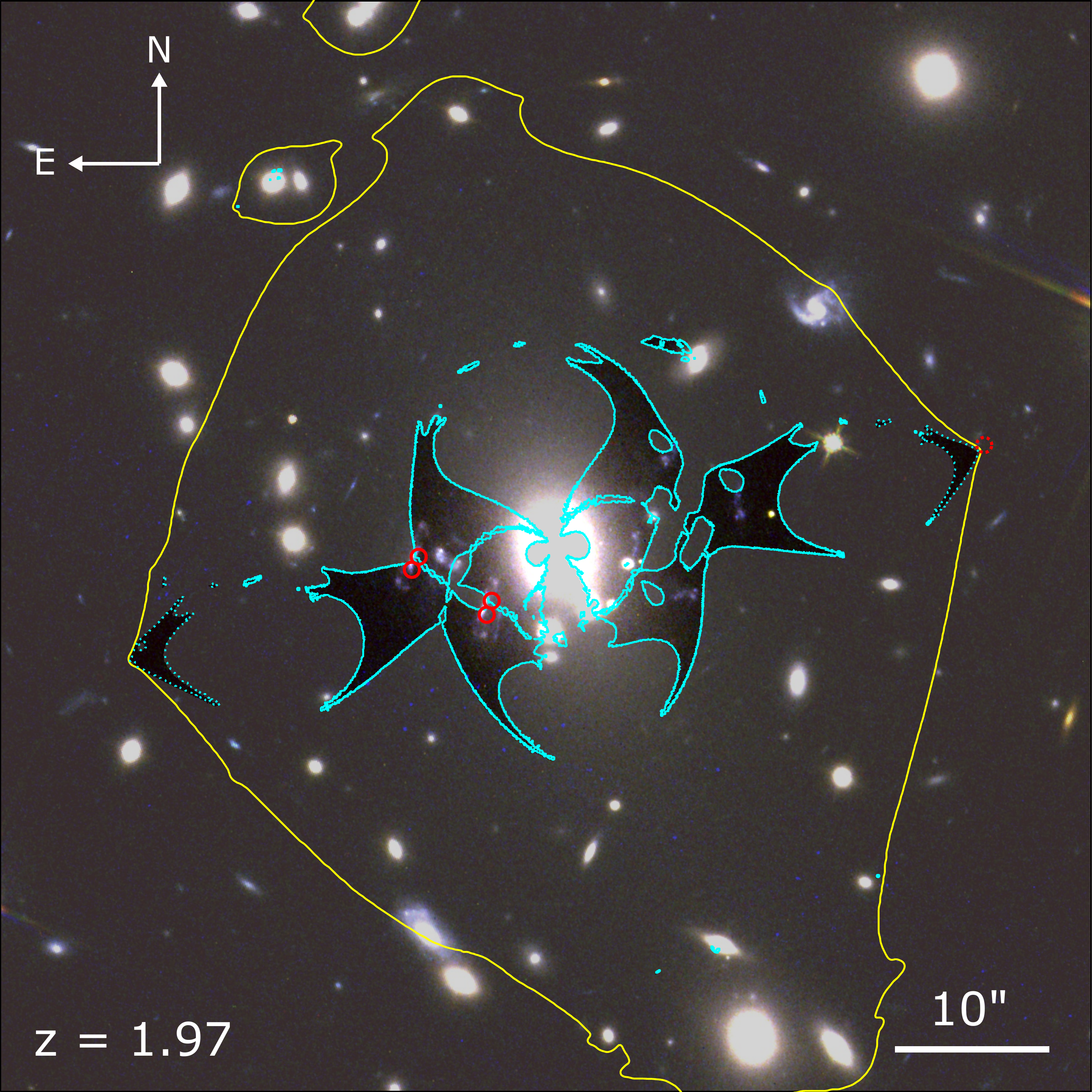}\hfill
      \includegraphics[width=0.32\textwidth]{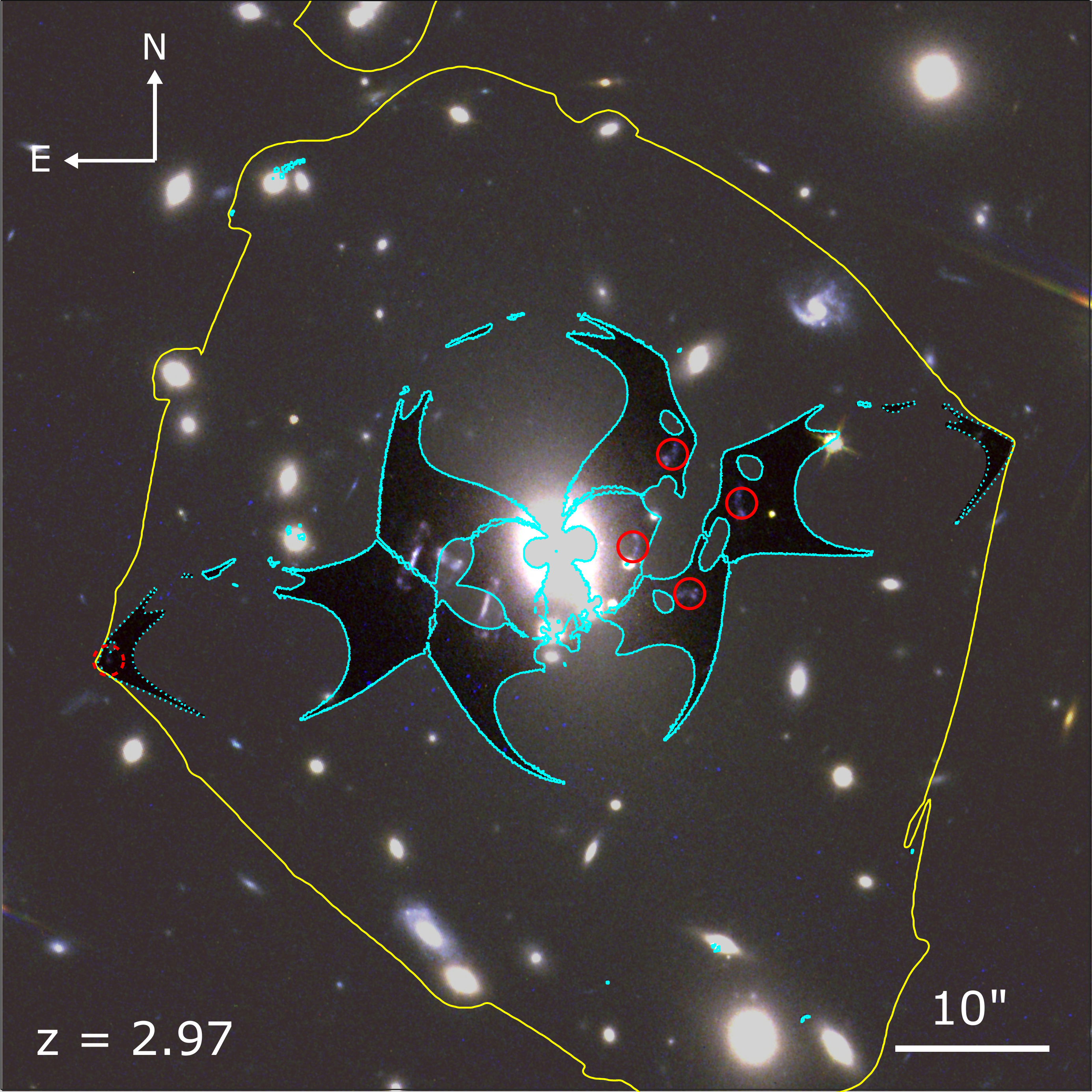}\hfill
      \includegraphics[width=0.32\textwidth]{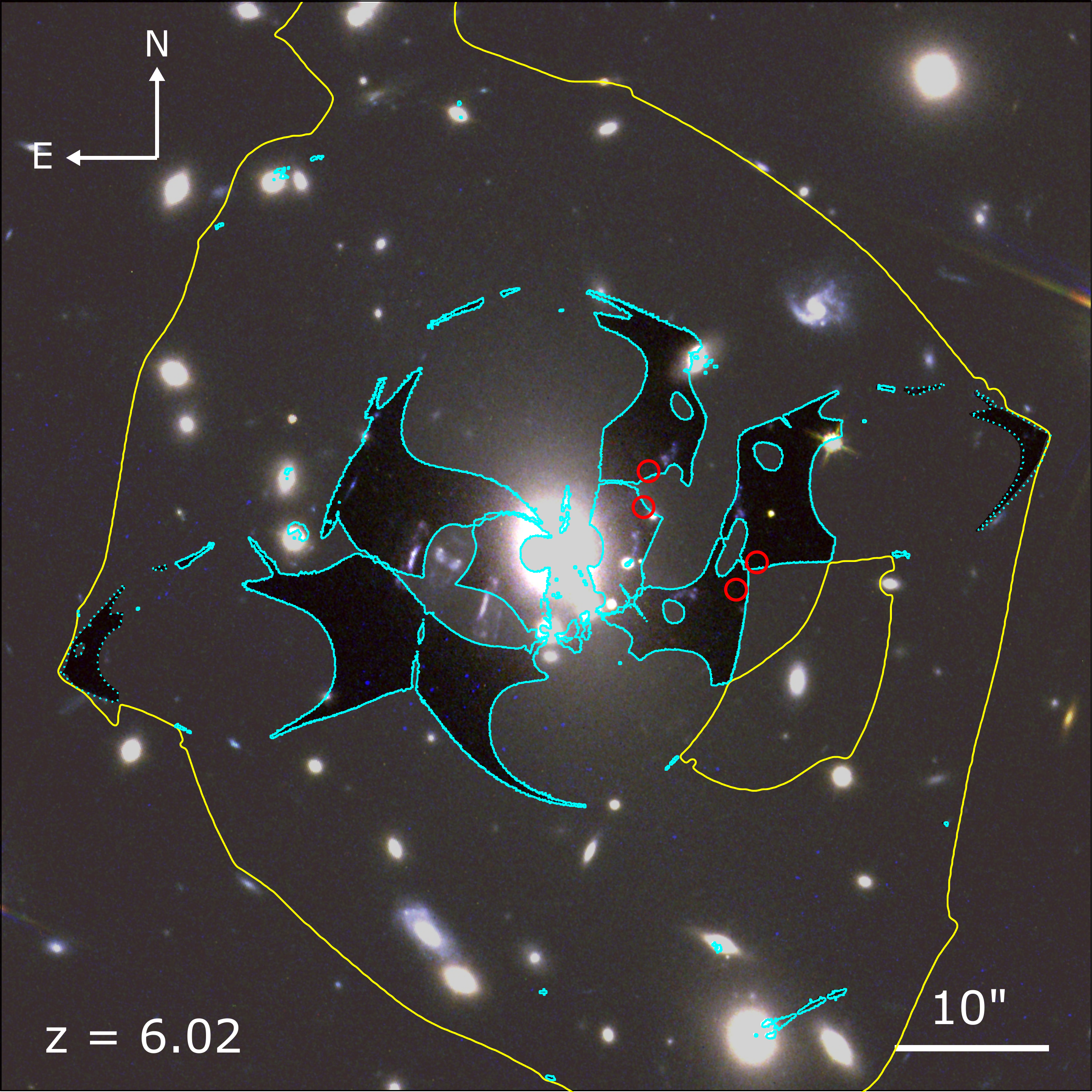}
      \caption{Exotic regions of RXJ0437+00 computed at source redshifts 1.97, 2.97, and 6.02 (blue), highlighting known HU exotic 4-image systems (red circles) and their predicted counter-images (dashed red circles). The dashed blue regions indicate the counter-image areas of the exotic regions, i.e. the corresponding counter-image space in the source plane. The yellow contour outlines the multiple-image region of the cluster at each of those redshifts. For the right-most panel ($z=6.02$), the fifth counter-image is not identified in the observations due to its faintness.
      }

      \label{masques_RXJ}
    \end{figure*}

\subsection{Definition of the exotic region}
\label{sec:exoticRegion}

    \begin{figure}[!ht]
      \centerline{\includegraphics[width=\columnwidth]{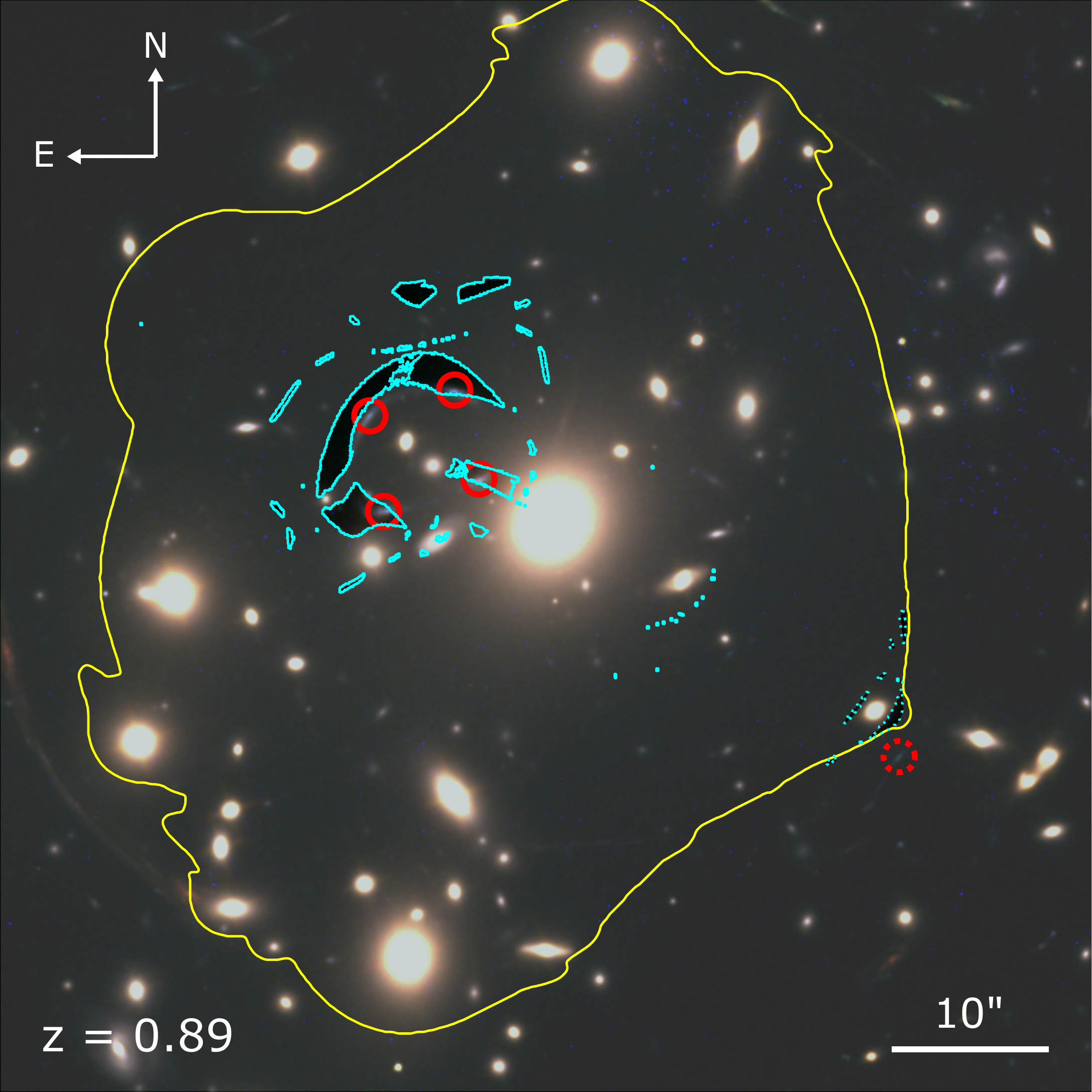}}
      \caption{Exotic regions of A1703 computed at source redshift $0.89$ (blue) overlaid on known HU exotic configurations (red). The dashed regions indicate the counter-image space associated with the exotic configurations, while the dashed lines mark the measured counter-image position. Small offsets between the predicted regions and the observed HU locations are likely to arise from the RMS positional uncertainty in the image plane induced by the Lenstool model. The yellow contour outlines the multiple-image region of the cluster at this redshift.}
      \label{masque_a1703}
    \end{figure}

    Identifying the exact cusp-exchange points is a necessary step, but it does not by itself characterise the predisposition of a galaxy cluster to produce HU-like exotic images. Strictly speaking, a hyperbolic umbilic exists only at a unique point in the source plane for a given redshift. In practice, however, sources located in the immediate vicinity of this point can still generate image configurations that remain highly magnified, nearly isotropic, and observationally distinctive. It is therefore natural to define, for each redshift, a finite region around the cusp-exchange point within which HU-like configurations can be produced.

    The construction of these regions relies on the generic properties of HU image configurations. In the canonical picture, a HU is associated with a compact quadruplet formed in the immediate vicinity of the critical curves, where the magnification is large, together with an additional, spatially separated counter-image. In catastrophe theory language, the quadruplet carries the local HU signature, while the counter-image is often treated as the additional image required by the global lens mapping rather than being part of the local catastrophe itself \citep[e.g.][]{meena2023exotic,schneider1992}. In idealised mappings, image parities provide an additional diagnostic, with the compact set displaying the expected alternation of parities across the local critical structure. In practice, in cluster lenses with discretised maps and frequent $n>5$ multiplicities, parities can become ambiguous to assign automatically for a large sample, especially when secondary weakly magnified images are present.

    In practice, we scanned values of \(D_{ls}/D_s\) from 0 to 1 in steps of 0.025 and computed, at each step, an image multiplicity map in the image plane using \textsc{Lenstool}. Regions with an image multiplicity greater than or equal to five were then projected pixel by pixel onto the source plane. The corresponding image systems were recovered by ray-tracing these source positions back to the image plane, providing both image positions and amplification values. These systems were subsequently filtered through the following procedure, designed to ensure a robust and unambiguous selection of HU-like configurations.

    \begin{figure*}
    \sidecaption
    \includegraphics[width=12cm]{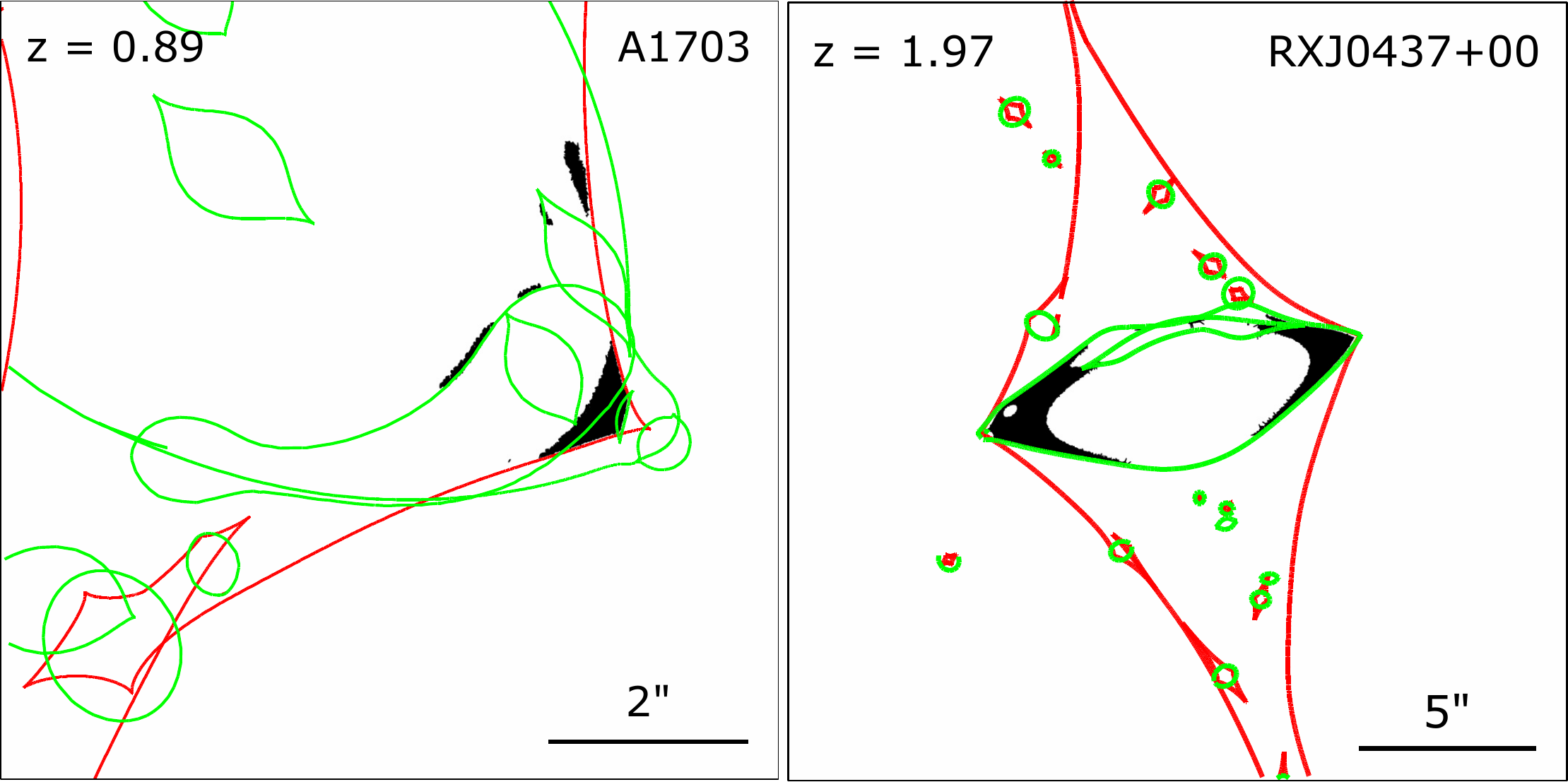}
    \caption{Exotic region in the source plane (black-coloured regions) for A1703 and RXJ0437+00 at $z = 0.89$ and $1.97$ respectively. Tangential and radial caustic lines are shown in red and green respectively.}
    \label{A3_source}
    \end{figure*}
    
    The first step was the identification of HU-eligible configurations. For each image system, all pairwise distances between images in the image plane were computed. The $n-1$ largest distances (where $n$ is the total number of images) were then selected. If the selected distances all involved a common image, this image was maximally separated from the others and was identified as an isolated image. The isolated image is then required to be the least magnified image of the system. This condition follows directly from the HU geometry: the compact HU-quadruplet lies close to the critical curves and is therefore strongly magnified, while the associated counter-image lies farther from the critical curves and remains weakly magnified. Systems that do not satisfy this hierarchy were rejected. In configurations with more than five images, secondary weakly magnified images may be present; in this case, these images were iteratively removed and the procedure was repeated as long as at least five images remained.

    This amplification hierarchy is consistent with the asymptotic HU regime, in which the characteristic HU quadruplet lies arbitrarily close to the critical curve and is therefore strongly magnified, while the associated counter-image remains farther from the critical structure and is only weakly magnified. This behaviour is captured by the HU proximity parameter $R_{\rm HU}$ introduced by \citet{meena2023exotic}, which measures how closely a given configuration approaches the ideal HU magnification relation: in the idealised limit, $R_{\rm HU}$ vanishes exactly at the HU point \citep{aazami2009universal}. 

    In practice, $R_{\rm HU}$ shows scatter in realistic cluster models and at finite numerical resolution, making a strict threshold model- and resolution-dependent. This motivates the introduction of a new identification criterion in this work.

    The HU proximity parameter and image parities are then used as a posteriori consistency checks rather than as primary selection criteria. For each retained configuration, we manually verified that the HU quadruplet exhibited the expected parity pattern $(+,-,+,-)$ using the \texttt{Lenstool} output files, which also provide access to the corresponding $R_{\rm HU}$ values. This approach ensured consistency with HU expectations while retaining the flexibility required to explore complex cluster lenses.
 
    The second step consisted in a cross-checking with HU-points. Each HU point corresponded to an exact source redshift, whereas the multiplicity maps were constructed on a discrete sampling of \(D_{ls}/D_s\). Each HU point was therefore associated with the nearest sampled redshift. For each retained configuration, we constructed the convex polygon defined by the images excluding the counter-image and tested whether the HU point lies inside this polygon. The configuration was kept only if this condition was satisfied at the sampled redshift closest to the exact HU redshift, ensuring a consistent association between HU points and image systems.
  
    The last step was a recursive extension to neighboring redshifts. Although a true HU exists only at a single source redshift, HU-like configurations persist over a narrow redshift interval. Once a configuration was validated at a given sampled redshift, the corresponding systems at adjacent redshift steps were also retained. This step ensures that the exotic region captures not only the mathematically exact HU but also the range of redshifts producing nearly isotropic, observationally distinct HU-like images.

    After this filtering, the exotic region in the source plane was defined as the collection of all source positions associated with the retained systems. This process provides a robust mapping of the areas where HU-like exotic image formation is possible, while preserving consistency with the exact redshift of each HU point and the amplification values of the selected sources.

    \subsection{Validation}
    \label{sec:validation}

    To validate our approach, we applied it to well-constrained cluster models that have been previously introduced as hosting HU exotic configurations. In these tests, we employed convergence and shear maps covering $200 \times 200$ arcsec, large enough to encompass each cluster's main strong-lensing region, and sampled at 0.066 arcsec per pixel. In particular, we examined RXJ0437.1+0043 (RXJ0437+00; $z=0.285$, \citealp{Lagattuta2023}) and Abell 1703 (A1703; $z=0.28$, \citealp{limousin2008,Richard2009}), both renowned for their visually recognised HU exotic images. For RXJ0437+00, high-quality MUSE, Keck/MOSFIRE, and HST imaging have revealed three exotic systems at source redshifts $\boldsymbol{z=1.97}$, $2.97$, and $6.02$. Similarly, A1703, a massive, X-ray-luminous cluster observed with HST (ACS, NICMOS) and from the ground with Subaru and Keck/LRIS, features a central ring of four images at $z=0.89$. In Figures~\ref{masques_RXJ} and~\ref{masque_a1703}, overlaying the computed exotic regions in blue onto the HST mosaics shows that these regions align well with the observed HU exotic configurations circled in red, in particular for the main quadruplet. We note, however, that the predicted location of the fifth image shows noticeable deviations in some cases, occasionally lying outside the nominal multiple-image region. This image, typically faint and less constrained by the central multiple-image configuration, likely makes its predicted position more sensitive to the details of the mass model and to the extrapolation of the lens potential. Overall, the agreement remains good for the main quadruplet, supporting the robustness of our method in identifying the regions where HU exotic configurations are expected to form.

    We further verified the consistency of our method by examining the source-plane morphology of exotic regions for two confirmed HU systems: A1703 at source redshift \(z = 0.89\) and RXJ0437+00 at \(z = 1.97\). Figure~\ref{A3_source} shows the exotic areas identified in the source plane at these redshifts. In both cases, the regions are remarkably compact ($0.70\,\mathrm{arcsec}^2$ for A1703 and $5.13\,\mathrm{arcsec}^2$ for RXJ0437+00), illustrating how spatially confined these regions are and how significant it is when a background source lies within them. When radial and tangential caustics are overlaid, the exotic regions align closely with the narrow exchange zone bounded by the two caustics, fulfilling the cusp-exchange criterion required for HU configurations.  

    In detail, several features deserve attention. First, while HU regions are generally expected to cluster around the cusp-exchange point of the radial and tangential caustics, this behaviour is not universal. In A1703, for instance, the cusp tip itself is not flagged as exotic, in contrast to RXJ0437+00 where the HU region remains tightly concentrated around the expected cusp-exchange zone, illustrating the diversity of HU morphologies across clusters. When such offsets occur, they may reflect the sensitivity of HU configurations to higher-order derivatives of the lensing potential, which determine the local structure of the magnification matrix and can shift the region where the HU criteria are satisfied relative to the geometrical cusp tip. In addition, asymmetries or small-scale structures in the mass distribution (e.g. cluster member galaxies or subhalos) may locally distort and displace the effective HU region from the nominal cusp-exchange point.

    Second, HU regions often display a composite structure, with a dominant and well-defined patch associated with the main cusp-exchange configuration, accompanied by multiple smaller satellite patches distributed in its vicinity. Inspecting and simulating the image configurations produced in these satellite regions shows that they correspond to HU-like solutions according to our selection criteria. Such patches are frequently associated with relatively simple local potentials, such as individual galaxy-scale halos embedded in the cluster environment, and naturally lead to localised agglomerations of HU points in the source plane, as illustrated in Fig.~\ref{a3_schema_a1703}. 

    In addition, the use of discretised lensing maps and a finite redshift sampling may lead the detection procedure to occasionally select secondary patches that are not associated with genuine cusp-exchange configurations. More generally, small-scale caustic features, particularly those related to galaxy-galaxy lensing, are highly sensitive to modelling choices (e.g. parametrisation of the lens potential, treatment of core radii, or cusp parameters). This explains why HU patches connected to these small-scale caustics, often located farther from the cluster centre, can differ between models.

    Together, these considerations underline that HU regions are not isolated singular locations but extended patches, composed of a robust central component and a surrounding population of satellite HU points. While their precise morphology depends sensitively on the detailed caustic structure, the overall robustness of our identification method remains unaffected.

    In the following section, we apply this detection method to a sample of galaxy clusters in order to explore the implications regarding the conditions that favour HU exotic image formation.

    \section{Application to massive lensing clusters}
    \label{sec:application}

    Now that we have validated the definition of HU exotic regions, we can use their identification to locate, in the image plane, where such configurations are most likely to appear for a fixed source redshift, providing a concrete visual guide for any subsequent inspection. At the same time, it is possible to quantify the predisposition of galaxy clusters to form HU configurations. To do so, we computed for each cluster the surface area of the exotic region in the source plane, \(\Omega(z)\), at each source redshift value. A larger exotic area at a given redshift implies a greater probability for a background galaxy to fall within this region, thus increasing the likelihood of HU image formation.

    To elevate this redshift-dependent information into a more generic, interpretable indicator, we defined the total exotic comoving volume \(V\). We computed it by summing \(\Omega(z)\) weighted by the differential comoving volume element \(\mathrm{d}V_{\rm com}(z)\) across redshift bins. The resulting volume \(V\) thus captures, from the source point of view, the full space within which HU configurations can form. Clusters with larger \(V\) are therefore more likely to produce exotic images.

    We present here the results obtained on a sample of galaxy clusters spanning a variety of physical properties (lensing strength, morphology, lens redshift) using maps covering 200 x 200 arcsec at 0.066 arcsec per pixel in the image plane. 

    \begin{figure}[!ht]
      \centerline{
      \includegraphics[width=\columnwidth]{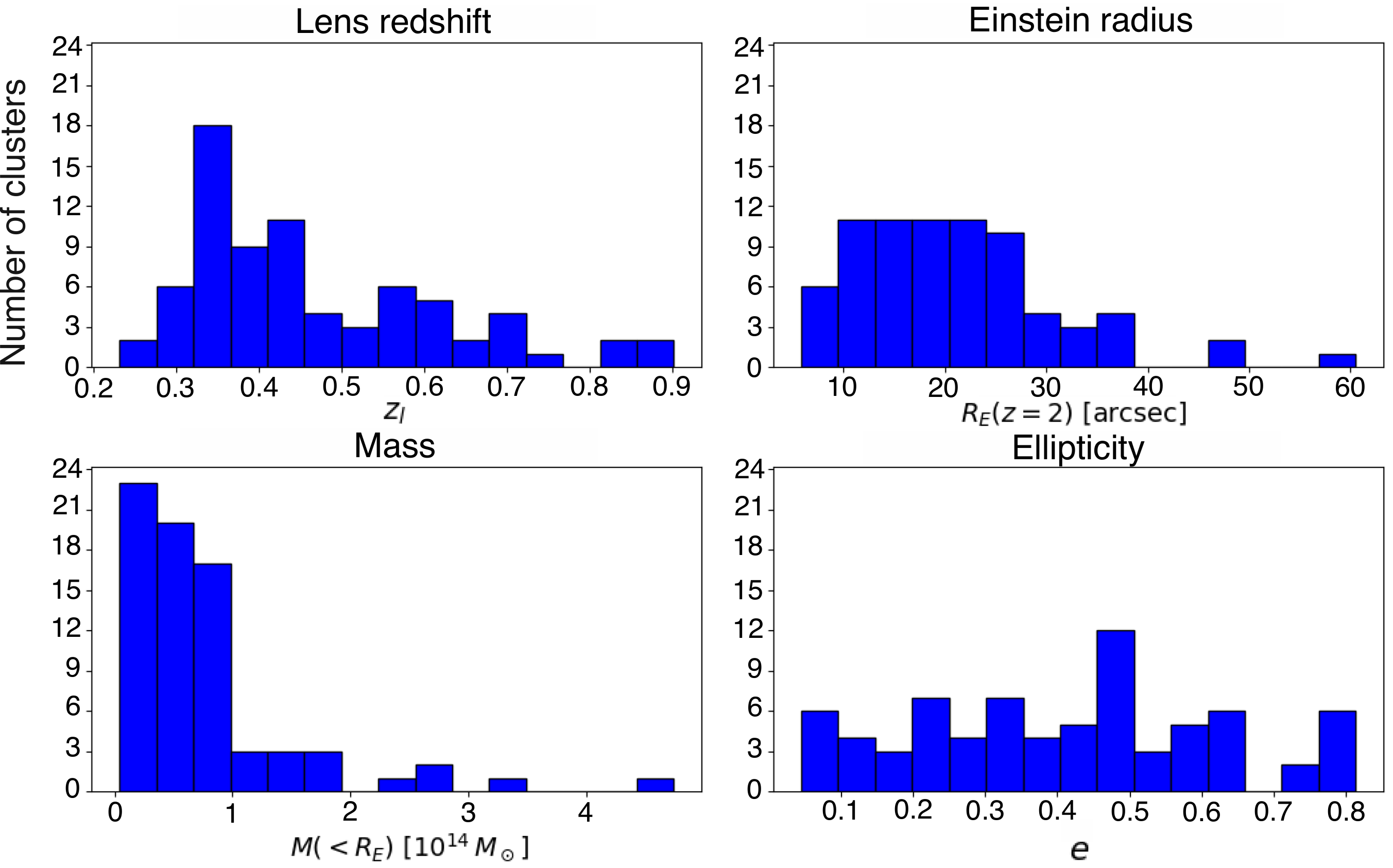}}
      \caption{Distribution of characteristics for our full sample of galaxy clusters: lens redshift \(z_l\), Einstein radius \(R_E\), enclosed mass within the Einstein radius \(M(<R_E)\), and ellipticity \(e\)}
      \label{histo_morpho}
    \end{figure}

\subsection{Description of the sample}
\label{sec:sample}

    We gather for this study a large sample of massive clusters with available strong-lensing models used previously in the literature (see Table~\ref{tab:complete}). We require these clusters to be modelled in a similar manner, using the Lenstool software and a parametric description of the mass distribution based either on dPIE (double Pseudo Isothermal Elliptical, \citealt{eliasdottir2007}) or NFW \citep{navarro1997} mass potentials. We also require all models to include at least one spectroscopic confirmation for a multiply imaged system used as a strong-lensing constraint, in order to guarantee a minimum quality of the model. The clusters in our sample are both massive and feature highly concentrated cores, as reflected by the distribution of their strong-lensing properties (Einstein radius \(R_{\rm E}\), enclosed mass \(M(<R_{\rm E})\), and ellipticity \(e\); see Figure~\ref{histo_morpho}).

    \begin{figure*}[!ht]
      \centering
      \includegraphics[width=\textwidth]{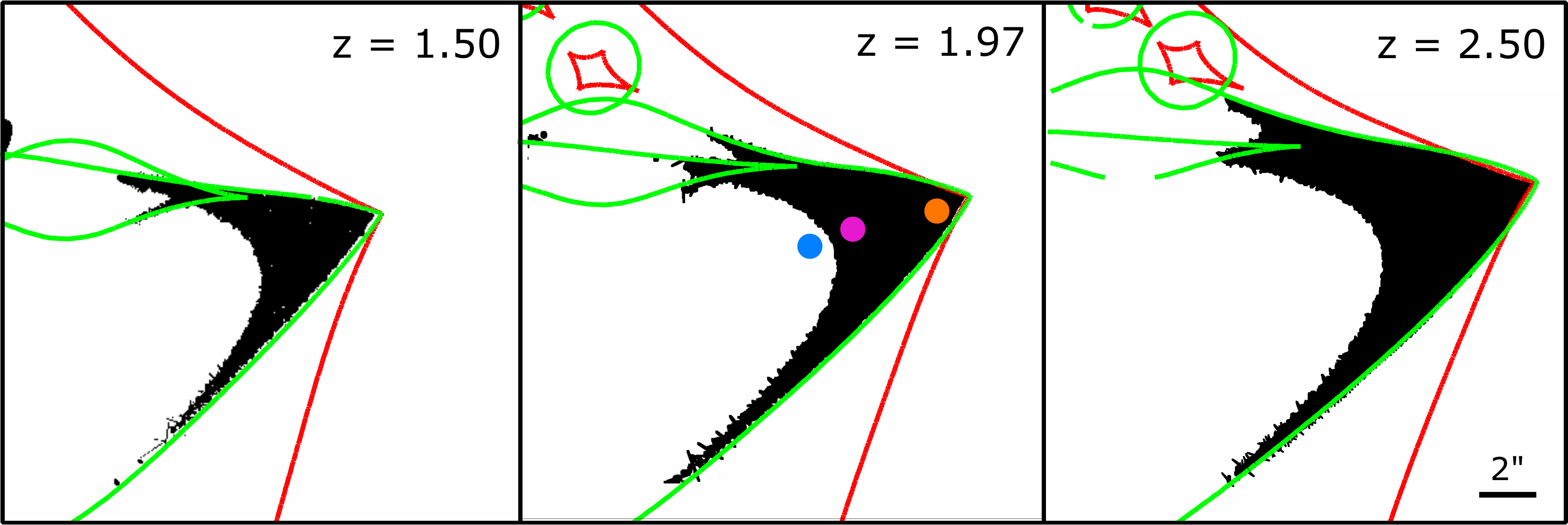}
      \caption{Evolution of tangential and radial caustic lines (red and green) with source redshift for RXJ0437+00. The scale is the same for the three panels computed for z=1.5, z=1.97 (the measured redshift of a confirmed HU exotic system) and z=2.5. Artificial point sources (blue, pink and orange) are placed in and out the exotic region at z=1.97.}
      \label{evol_caustic_image}
    \end{figure*}

    The majority of the models come from large samples of strong-lensing clusters observed as part of the MACS \citep{Ebeling_2007, repp2018}, eMACS \citep{ebeling2025}, LoCuSS \citep{richard2010}, Frontier Fields \citep{Lotz2017}, CLASH \citep{postman2012cluster}, RELICS \citep{coe2019}, and MUSE ATLAS \citep{Richard2021, claeyssens2022} efforts. The use of a large variety of cluster samples ensures significant statistics together with a broad range in cluster properties. In particular, the cluster redshifts range from $z_l=0.2$ to $z_l=0.9$. We do note a possible drawback of our sample, namely the heterogeneous selection functions, which rely on different criteria such as X-ray luminosity, mass concentration, or the size of the strong-lensing region. However, since our main objective is to investigate the diversity of HU exotic regions and their relation to cluster properties, this heterogeneity should not significantly affect our results, provided that the clusters span a wide range of morphologies and central mass concentrations.

    To consistently characterise the size and shape of each lens, we analyse all convergence maps rescaled to a fiducial source redshift \(z_s=2\). This choice is physically motivated: \(z\!\sim\!2\) coincides with the peak of the cosmic star-formation rate density, providing abundant bright, clumpy sources behind clusters, and it lies near typical redshifts of strongly lensed arcs, yielding representative \(D_{LS}/D_S\) values for lenses at \(z_l\approx0.3{-}0.6\) without extrapolating to very low or very high \(z_s\) \citep{madau2014cosmic, jauzac2015hubble}. The Einstein radius \(R_{\rm E}\) is defined from the largest critical curve: its enclosed area \(A_{\rm crit}\) is converted into an effective circular radius,
    \[
    R_{\rm E} = \sqrt{\frac{A_{\rm crit}}{\pi}},
    \]
    following the approach of \citet{meneghetti2023}.

    The ellipticity is measured from the \(\kappa\)-isocontour enclosing the largest region for which the average convergence satisfies \(\langle \kappa \rangle \simeq 1\). In the circular case, this threshold corresponds exactly to the Einstein radius, since
    \[
    \bar{\kappa}(<\theta_E) = \frac{1}{\pi \theta_E^2} \int_0^{\theta_E} 2\pi \theta\, \kappa(\theta)\, d\theta = 1.
    \]
    For more general mass distributions, this condition still provides a meaningful approximation of the strong-lensing boundary. By selecting the largest region satisfying \(\langle \kappa \rangle \approx 1\), we trace the effective zone where multiple imaging occurs, while avoiding sensitivity to small-scale features or noise near the core.

    An ellipse is fitted to this region via least-squares minimisation in image space. The ellipticity is defined as
    \[
    e = 1 - \frac{b}{a},
    \]
    where \(a\) and \(b\) are the semi-major and semi-minor axes of the best-fit ellipse.

    Uncertainties on both $R_{\rm E}$ and $e$ are estimated through bootstrap resampling of the isocontour points. By repeatedly subsampling the original contour and recalculating the associated quantity, we obtain a distribution of values whose standard deviation defines the statistical error. These Einstein radius and enclosed mass estimates were cross-validated against published values for the corresponding lens models when available, in particular for the recently released eMACS sample \citep{ebeling2025}, showing good agreement with our measurements. The resulting parameter uncertainties are then used in a first-order approximation to derive the error bars shown in Figure~\ref{fig:ccv_vs_params} (Section~\ref{sec:parameters}).

    \begin{figure*}[!bt]
      \centering
      \includegraphics[width=\textwidth]{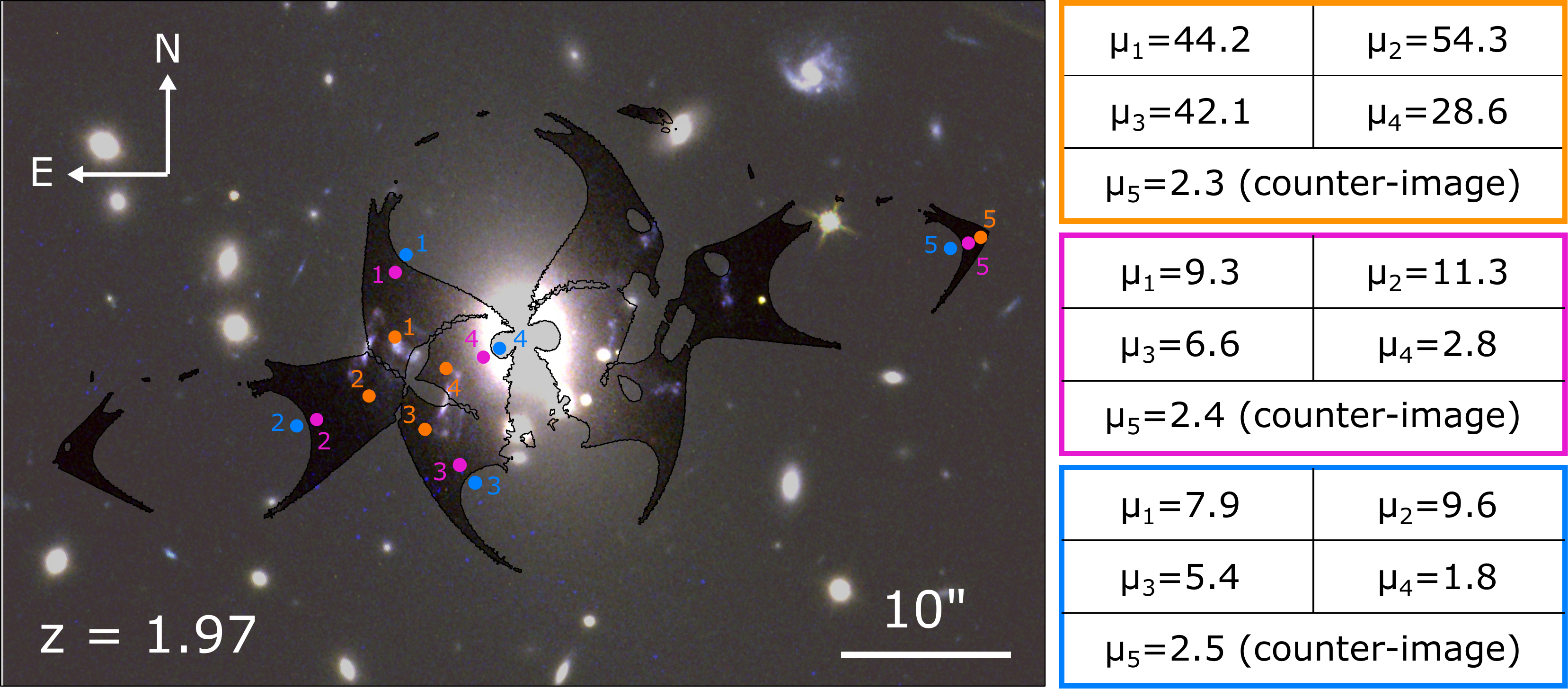}
      \caption{Image-plane projections of three point sources at $z=1.97$ (left), colour-coded as in Figure~\ref{evol_caustic_image}, with their magnification factors shown in the coloured table (right).}
      \label{proj}
    \end{figure*}

    \begin{figure*}
    \sidecaption
    \includegraphics[width=12cm]{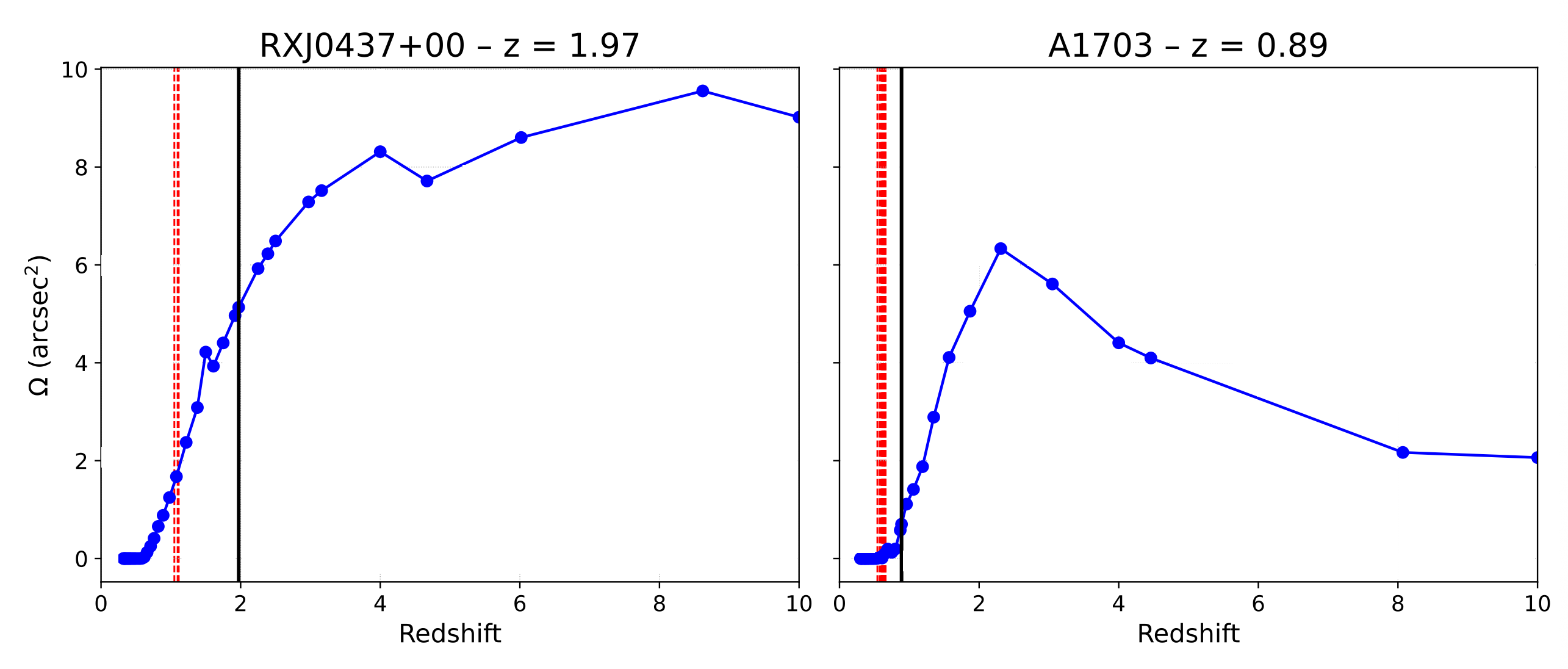}
    \caption{Exotic area \(\Omega\) versus source redshift for RXJ0437+00 (left) and A1703 (right). In each panel, the blue curve traces the measured area from $z=0$ to $z=10$. Dashed red vertical lines mark the redshifts at which our algorithm predicts HU-exotic regions within which the confirmed systems were found.}
    \label{surface_vs_z}
    \end{figure*}

    In total, we include 74 clusters satisfying our criteria, which we report in Table \ref{tab:complete} together with their Einstein radius, ellipticity, enclosed mass, lens redshift, as well as the references for the previous studies. 
    
    Figure~\ref{histo_morpho} presents histograms of \(R_{\rm E}\), \(e\), \(M(<R_{\rm E})\), and \(z_l\) for this cluster sample. The distributions span broad ranges in size, shape, mass, and lens redshift and highlights the diverse physical properties captured by our selection.

\subsection{Exotic region area and its evolution with redshift}

    HU exotic regions typically lie near a cusp-exchange point, both in sky coordinates and along the redshift axis. While their morphological significance was previously discussed in Section~\ref{sec:validation}, it is now important to emphasise that the evolution of these regions with source redshift directly impacts their observability. 
    
    To illustrate this, Figure~\ref{evol_caustic_image} tracks the positions of the radial and tangential cusps for RXJ0437+00 at three representative redshifts (one below, one at, and one above the expected exchange point). The overlaid curves clearly reveal the cusp exchange between the tangential and radial caustic line as we pass the redshift $z = 1.97$ corresponding to the redshift of the real HU exotic system. Around this critical redshift value, there is only a slow evolution of the exotic region on sky, which motivates the need to explore a broad redshift range to estimate the full volume associated with HU exotic configurations.

    Figure~\ref{proj} shows the projection in the image plane of the three coloured source points at $z = 1.97$ identified in Figure~\ref{evol_caustic_image}. The orange source lies well inside the exotic region near the cusp-exchange point, the pink source sits just inside the region's boundary on the opposite side, and the blue source lies just outside, close to the pink location. The adjacent table lists the magnification factors for all images produced by each source. The orange configuration exhibits the characteristic HU exotic signature, with its counter-image still having by far the lowest magnification, markedly separated from the significantly higher values of the other images in the system. In contrast, the blue source yields a classical quadruple-image system, since its counter-image is not the least amplified. The orange source not only satisfies the HU criteria but also achieves substantially higher magnifications.

    Figure~\ref{surface_vs_z} presents the evolution of the exotic region area \(\Omega(z)\) as a function of source redshift for RXJ0437+00 and A1703. The A1703 profile exhibits a pronounced peak, with \(\Omega\) rising to \(\sim6.3\,\mathrm{arcsec}^2\) at \(z\approx2.35\) before declining, whereas RXJ0437+00 follows a nearly monotonic trend across the same redshift range. We also note that the HU redshift predicted by our algorithm for the regions containing the confirmed HU configurations (dashed red lines) lies below the actual redshifts of those observed systems (solid black lines). From an observational standpoint, two implications follow. First, very deep surveys are essential to probe the widest possible redshift windows, especially since we now know that the peak of the HU exotic area can lie beyond the redshift of any currently confirmed system. Second, the variety of trends seen in \(\Omega(z)\) profile shapes, driven by variations in cluster structure, suggests tailored observing strategies: for lenses like A1703, targeting redshifts near the predicted peak maximises the chance of detecting extended exotic regions; for systems similar to RXJ0437+00, probing the highest feasible source redshifts yields the largest exotic areas. In both cases, recognising the characteristic shape of the \(\Omega(z)\) profile informs the optimal design of future HU-exotic imaging campaigns.

    Another key point concerns the quantification of the degree of isotropy achieved within the HU regions identified across our cluster sample. To do so, we characterise the local anisotropic deformation through the eigenvalues of the Jacobian matrix of the lens equation, $\alpha$ and $\beta$, which directly control the principal stretching factors of lensed images \citep{schneider1992,bartelmann2001weak}. In this framework, isotropic amplification corresponds to comparable stretching along the two principal directions, while anisotropic configurations are characterised by a strong imbalance between $\alpha$ and $\beta$. We therefore define the dimensionless stretching ratio \[ q \equiv \left|\frac{\alpha}{\beta}\right|, \] for which $q \simeq 1$ indicates nearly isotropic amplification, whereas $q \ll 1$ or $q \gg 1$ corresponds to strongly anisotropic, shear-dominated deformations. Computing $q$ over the HU regions identified by our procedure from the full sample of HU, we obtain a median value of $1.62$, with a $16$-$84\%$ range of $0.65$-$3.17$. Within individual HU regions, larger intra-region variations are observed, primarily driven by pixels located very close to the critical curves where one eigenvalue approaches zero. However, the bulk of each HU region retains comparable stretching along the two principal directions, indicating that the amplification does not overwhelmingly favour a single axis and supporting the interpretation of HU configurations as efficient natural telescopes.

    This behaviour is also illustrated by the three example systems shown in Fig.~8. For the orange HU system, whose source lies close to the cusp-exchange point, we obtain a mean value $\langle q \rangle = 3.49$, with values ranging from $2.15$ to $5.16$. The pink HU system, located further away from the cusp-exchange point while still remaining inside the HU region, exhibits a slightly larger anisotropy with $\langle q \rangle = 3.63$ and a broader range of $0.37$-$8.21$. This increase is consistent with the larger distance from the cusp-exchange configuration in the source plane. Finally, the blue system located outside the HU region presents the strongest anisotropy, with $\langle q \rangle = 3.89$ and values spanning $0.24$-$8.59$. Both the larger average value and the broader dispersion therefore support the interpretation that amplification becomes increasingly anisotropic outside the HU regions.

    \subsection{Exotic region surface and exotic comoving volume}
    
    For each cluster, we derive two exotic surface measures and two comoving volume measures, computed over two source-redshift intervals: a conservative range (\(z_{\rm l}\le z\le4\)), corresponding to the regime of high completeness in multiple-image identification—particularly through Lyman-\(\alpha\) emitters detected with MUSE \citep{Richard2021}, and an extended one (\(z_{\rm l}\le z\le10\)), providing a uniform upper observational limit across clusters, reflecting the rarity of strongly lensed sources beyond \(z=10\). \(\Omega_{\max,\,z<4}\) is defined as the maximum exotic source-plane area attained within the (\(z_{\rm l}\le z\le4\)) redshift interval.
    
    Table~\ref{tab:complete} summarises \(\Omega_{\max,\,z<4}\), \(\Omega_{\max,\,z<10}\), \(V_{z<4}\), and \(V_{z<10}\) for each cluster.  

    \begin{figure}[!h]
      \centerline{\includegraphics[width=\columnwidth]{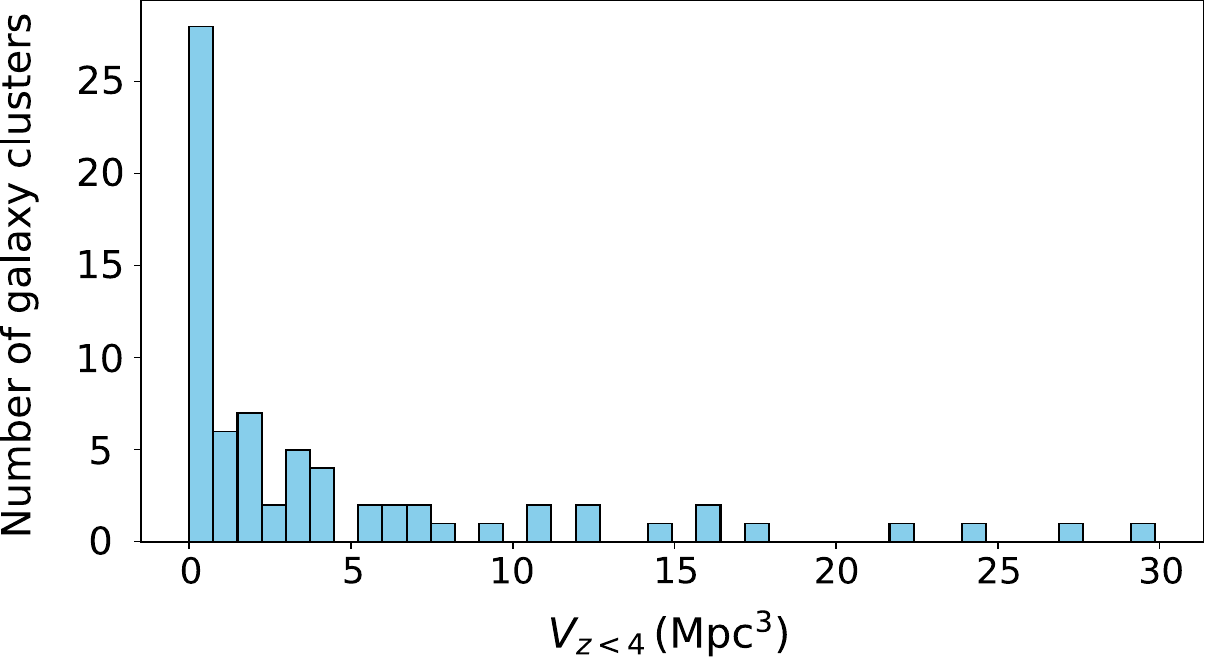}}
      \caption{\(V_{z<4}\) distribution for our sample.}
      \label{histo_volume}
    \end{figure}

    One of the first results from this analysis is that we directly notice that a significant number of clusters ($18 \%$) have \(V_{z<4} < 0.1\)\,Mpc\(^3\)  at this map resolution. Conversely, clusters such as RXJ0437+00 and A1703 stand out, with \(V_{z<4}\) equal to 15\,Mpc\(^3\) and 13\,Mpc\(^3\), respectively. Interestingly, we can see that several clusters show a large exotic volume per our criteria but have not yet been presented as hosting any HU exotic systems in the multiple systems used as strong-lensing constraints (e.g. AS1063, eMACS1527, \dots). This may be attributed either to the absence of a source within the predicted exotic volume and/or to insufficient observational depth. The latter is supported by the fact that the most promising clusters typically exhibit their maximum exotic area at source redshifts exceeding 5 (and therefore requiring deep, high-resolution images in the near infrared) while the majority of the clusters in the sample rely on HST images at optical wavelengths.

    \begin{figure}[!ht]
      \centerline{\includegraphics[width=\columnwidth]{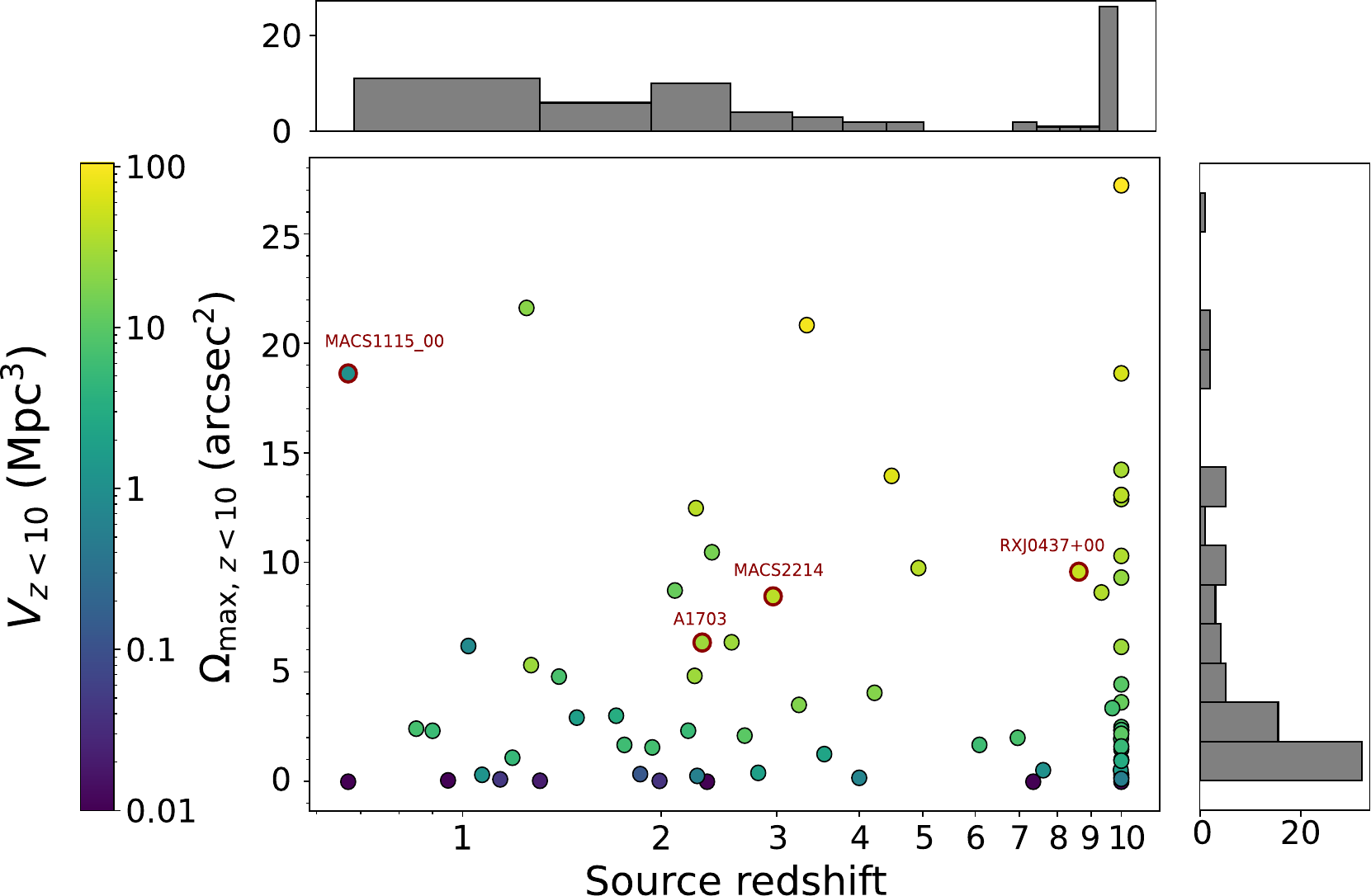}}
      \caption{Maximum exotic area \(\Omega_{\max,\,z<10}\) as a function of the source redshift. Point colour encodes the exotic comoving volume \(V_{z<10}\) for each cluster. Histograms display the distribution of points along the redshift and area axes.}
      \label{smax_vs_z}
    \end{figure}

    \begin{figure*}[!ht]
      \centerline{\includegraphics[width=\textwidth]{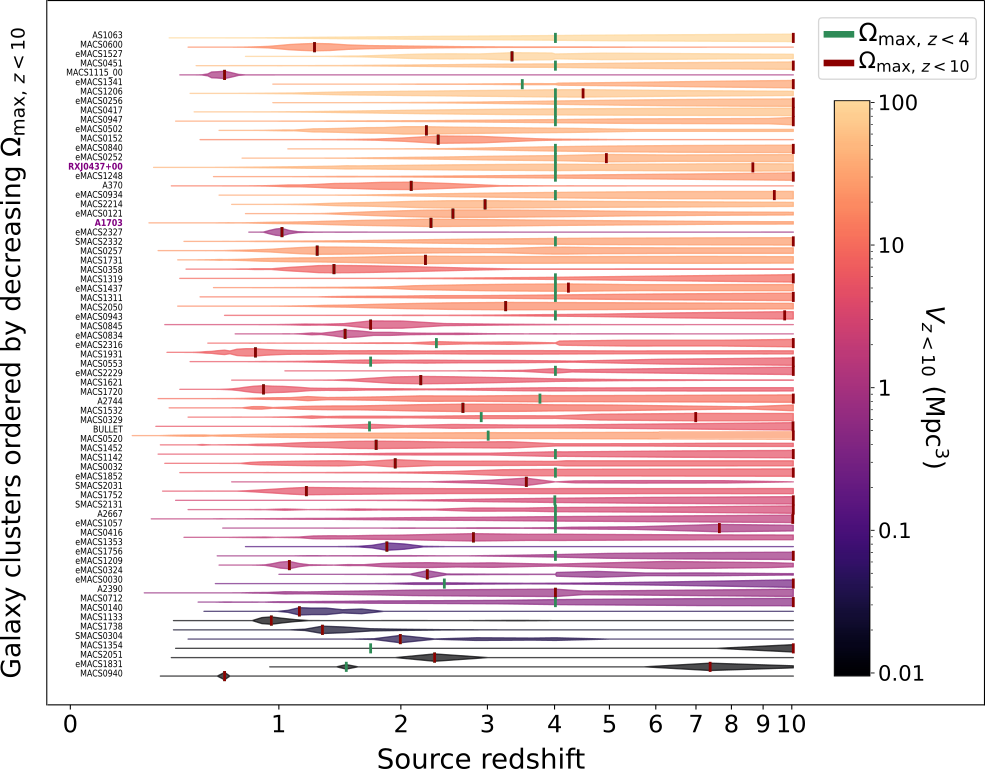}}
      \caption{Evolution of the exotic area as a function of source redshift. For each cluster, the envelope is normalised by its own maximum value over the explored redshift range. The colour of each envelope encodes the exotic comoving volume up to $z=10$. Green and red ticks indicate, for each cluster, the source redshifts at which the exotic area reaches its maximum when restricting the calculation to $z<4$ (\(\Omega_{\max,\,z<4}\)) and when extending it up to $z<10$ (\(\Omega_{\max,\,z<10}\)), respectively. Finally, clusters are ordered by decreasing \(\Omega_{\max,\,z<10}\) from the top to the bottom of the figure.}
      \label{synthese}
    \end{figure*}

    Figure~\ref{histo_volume} shows the \(V_{z<4}\) distribution for all clusters in our sample. With an average volume of approximately 4.59\,Mpc\(^3\)and more than 72\% of the clusters falling below this value, it is clear that clusters capable of producing HU exotic images are rare. This helps explain the low number of HU exotic observations reported in the literature and is further discussed in Section~\ref{sec:HU_number}. We illustrate the relationship between \(\Omega_{\max,\,z<10}\) and \(V_{z<10}\) in Figure~\ref{smax_vs_z}. We first note that, in principle, large exotic areas can occur at any source redshift, but to be effectively observed, a sufficient background galaxy density at that redshift is required: a sizable region does not guarantee that galaxies will actually lie within it (see Section~\ref{sec:HU_number}). We also observe that a larger \(\Omega_{\max,\,z<10}\) generally corresponds to a higher \(V_{z<10}\) value. However, there are notable exceptions. Some clusters (e.g. MACS2214, highlighted in Figure~\ref{smax_vs_z}) exhibit a high cumulative volume but a relatively small maximum exotic area, indicating small exotic surfaces spread over a broad redshift interval. Conversely, other clusters (e.g. MACS1115+0129, referred to as MACS1115\_00, also shown in Figure~\ref{smax_vs_z}) display a large maximum exotic area yet low cumulative volume, indicating that their exotic region is confined to a narrow redshift interval and/or occurs at low redshift, where the comoving volume per unit area is smaller. In both cases, detecting HU exotic systems becomes more challenging, either because deeper observations are required or the probability of a source lying within the exotic region is too low.

    Figure~\ref{synthese} synthesises these results, showing for each cluster the evolution of \(\Omega(z)\) up to a source redshift of \(z = 10\). Each envelope, normalised by \(\Omega_{\max,\,z<10}\), is colour-coded by the exotic comoving volume \(V_{z<10}\). As hinted in Figure~\ref{surface_vs_z}, a clear diversity emerges: some clusters exhibit a steadily increasing exotic region with redshift, while others show nonzero exotic areas only within specific redshift interval. Clusters with broad envelopes maintain significant exotic area across many redshifts, whereas those with high peaks, ranked at the top of the plot (clusters being ordered by decreasing \(\Omega_{\max,\,z<10}\)), maximise the chance of a background galaxy intersecting the region at a given redshift. Together, these two attributes yield a substantial exotic volume, thereby boosting the probability of hosting an exotic source within.

\begin{figure*}
    \sidecaption
    \includegraphics[width=12cm]{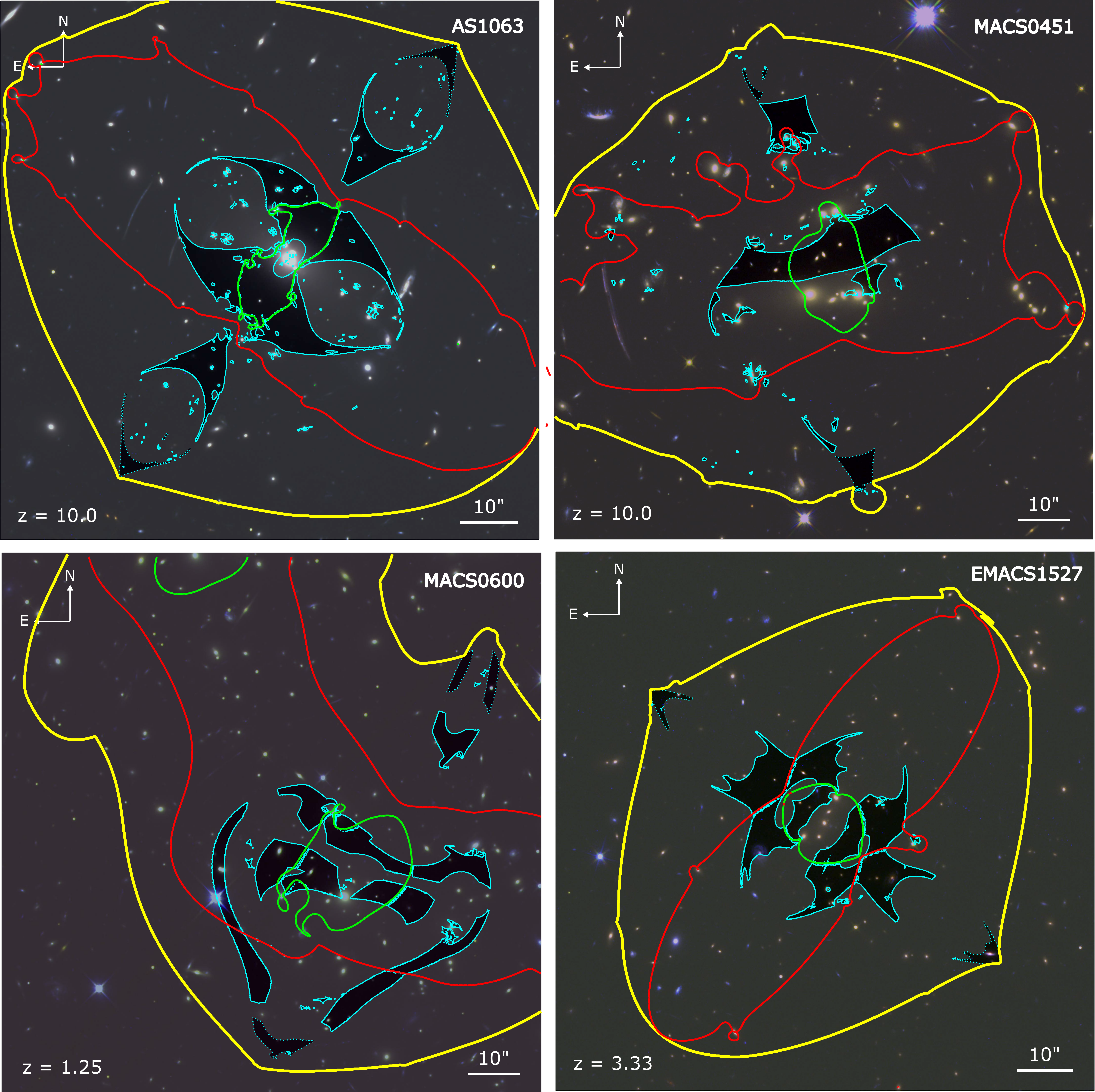}
    \caption{Exotic regions in the image plane (blue) for the four clusters with the highest \(V_{z<10}\) in our sample. In each panel, the source is placed at the redshift that maximises the exotic area \(\Omega\), and the yellow contour shows the corresponding multiple-image region, and the green and red contours show the tangential and radial critical lines respectively, all computed at this source redshift. Dashed lines indicate the observed counter-image positions.}
    \label{fig:exotic_mosaic}
    \end{figure*}

    Finally, Figure~\ref{fig:exotic_mosaic} displays the exotic regions in the image plane for the four clusters that exhibit the highest \(V_{z<10}\) in our sample. None of these clusters has yet yielded a confirmed HU exotic system in the literature. As discussed above, the exotic area can grow monotonically with increasing source redshift for some clusters such as AS1063 and MACS0451, implying that the deepest observations offer the greatest chance of detecting HU exotic images. In contrast, clusters MACS0600 and eMACS1527 reach their maximal exotic area at lower redshift and then decline, indicating a preferred specific redshift window for observing HU exotic configurations.

\section{Discussion}
\label{sec:discussions}

    The method we have presented and tested to estimate a cluster's predisposition to produce HU exotic configuration relies on a number of assumptions in particular on a perfect modelling of the mass distribution. We discuss here the robustness of the method and explore the relation between the exotic volume and some of the structural parameters of the cluster mass distributions. We also estimate the minimum number of HU configurations we can expect from the current data under conservative assumptions.

\subsection{Robustness of the exotic region calculation}

    To assess the robustness of the method presented in this work, we employed two complementary approaches to quantify potential sources of error.   
    
\begin{figure*}
    \sidecaption
    \includegraphics[width=12cm]{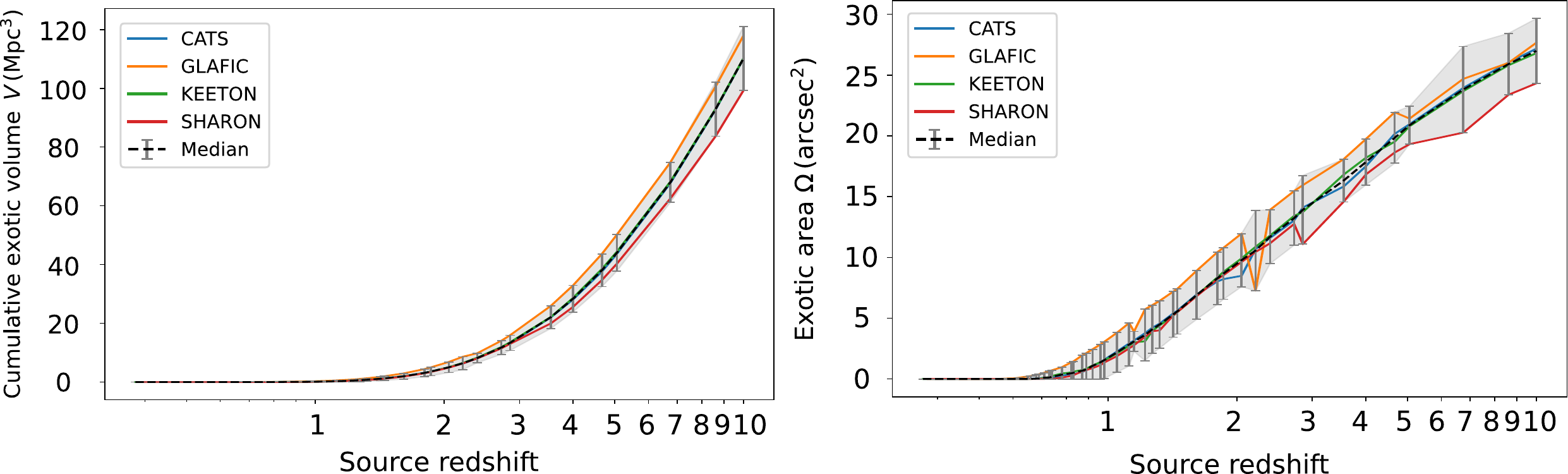}
    \caption{Systematic uncertainty in the exotic surface area (left) and the cumulative exotic comoving exotic volume (right) as functions of source redshift for AS1063. Results are shown for four independent lens models (CATS, GLAFIC, Keeton, and Sharon), and the shaded band at each redshift represents the maximum deviation from the median value.}
    \label{fig:systematics_error}
    \end{figure*}

    We first take advantage of the publicly available strong-lensing maps for the Frontier Field cluster Abell S1063,\footnote{\url{https://archive.stsci.edu/pub/hlsp/frontier/abells1063/models/}} which ranks among the clusters with the largest exotic volumes in our sample (\(V_{z<10} = 104.94\) Mpc\(^3\)). These maps, produced by Natarajan \& Kneib (CATs, \citealp{richard2014}), the GLAFIC team \citep{kawamata2018}, Sharon et al. \citep{johnson2014}, and Keeton et al. \citep{raney2020}, are all constrained by extensive sets of multiple images. We run all of these maps through \textsc{Lenstool} and our detection algorithm, so that the evolution of the exotic area and of the cumulative comoving exotic volume as functions of source redshift can be derived homogeneously (Figure~\ref{fig:systematics_error}). While we adopt the CATs model as our reference throughout this paper, this procedure allows us to estimate the systematic uncertainty arising from model choice by comparing the results obtained from each set of maps. The absolute error at each redshift is defined as the maximum deviation between the median value and all four models. Overall, this systematic uncertainty typically amounts to about $10\%$ for the exotic area and for the exotic comoving volume.

\begin{figure*}
    \sidecaption
    \includegraphics[width=12cm]{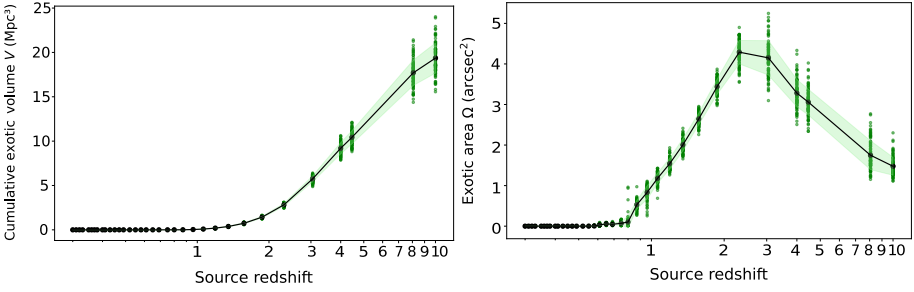}
    \caption{Intrinsic MCMC uncertainty in the exotic surface area (left) and the cumulative exotic comoving volume (right) as functions of source redshift for cluster A1703. Curves show the mean over 100 MCMC realisations, with the error bars indicating the $\pm1\sigma$ scatter at each redshift.}
    \label{fig:mcmc_error}
    \end{figure*}

    Second, recognising that most parametric models in this study were obtained via the Lenstool MCMC optimiser, we evaluated the intrinsic uncertainty of this process by selecting 100 model realisations drawn directly from the Lenstool MCMC chain for one of the cluster where we have a confirmed HU configuration: A1703. Applying our detection algorithm to each of those models (Figure~\ref{fig:mcmc_error}), we quantified the resulting uncertainty by computing the mean and standard deviation of the exotic area and cumulative exotic comobile volume at each redshift. This MCMC-driven error similarly grows with redshift, attaining characteristic values of $\sim 10\%$ for the surface and $\sim 9\%$ for the volume.

    Together, these analyses demonstrate that, within the family of parametric models considered here, the uncertainties on the exotic area and volume estimates are comparable to those affecting standard lensing observables, such as mass profiles, and remain relatively insensitive to modelling assumptions as long as the cluster is constrained by a sufficient number of multiple images. A complementary perspective is provided by the work of \citet{meena2021exotic}, who showed that free-form reconstructions yield systematically fewer point singularities and hence an order-of-magnitude reduction in the exotic cross-section (see their Fig.~3). This apparent tension illustrates that the impact of modelling choices on exotic lensing is non-negligible. Part of the difference may also stem from the methodology: in \citet{meena2021exotic}, cross-sections were derived within a comoving cylinder of radius 5 kpc integrated over a finite redshift interval to estimate exotic volumes (see their Sect. 4.3), whereas our analysis adopts an observationally motivated, redshift-dependent approach linking each theoretical HU point to a finite window of HU-like configurations. These distinctions highlight the need for caution when comparing across modelling families, but they do not undermine the robustness of our estimates within the parametric framework adopted in this work.

\subsection{Correlations between cluster parameters and exotic volume}
\label{sec:parameters}

    In order to assess which physical properties of galaxy clusters favour the formation of HU image configurations, we have extracted the following structural and morphological parameters for each lens in our sample and examined their correlations with the total exotic comoving volume \(V_{z<10}\). These parameters, along with their values for all clusters, are summarised in Table~\ref{tab:complete}:

    \begin{figure*}[!ht]
      \centering
      \includegraphics[width=\textwidth]{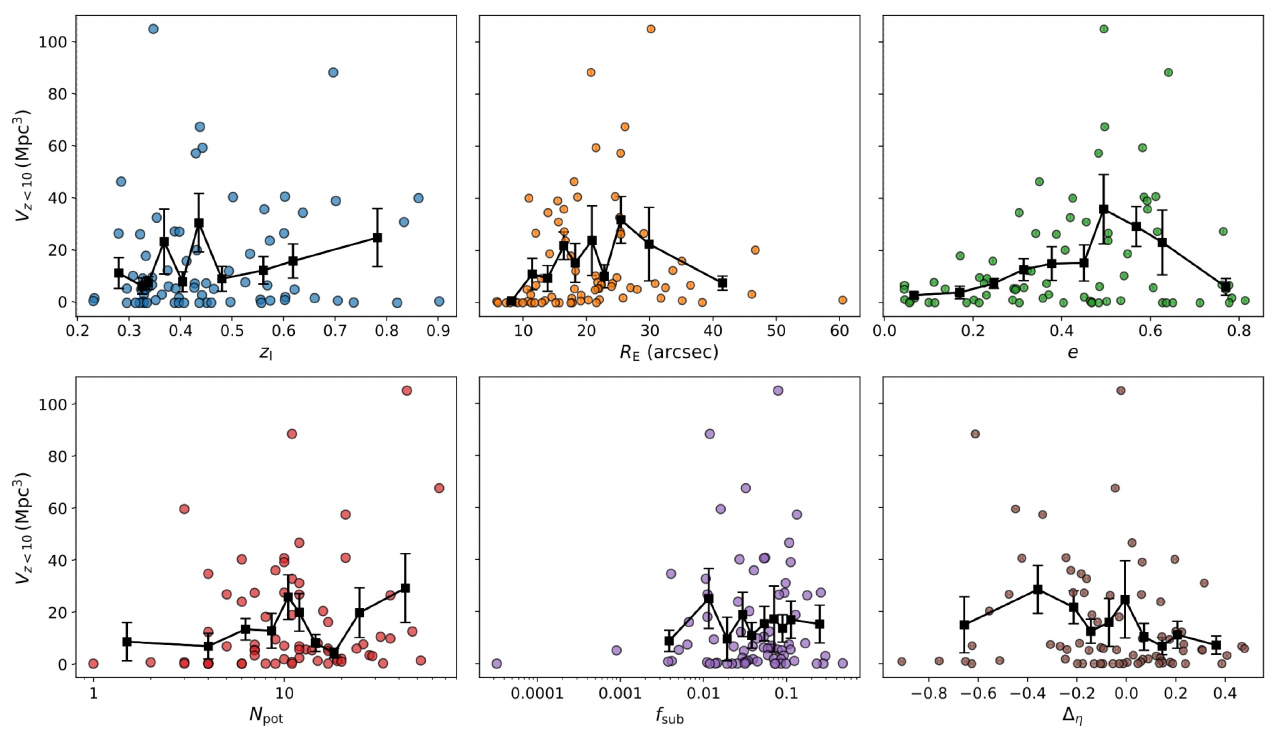}
      \caption{Exotic comoving volume \(V_{z<10}\) versus six individual cluster parameters. Coloured markers denote each cluster (error bars on parameters where available), while black squares connected by lines show the mean \(V_{z<10}\) in quantile bins of the horizontal axis parameter, with \(\pm1\,\mathrm{SE}\) error bars.}
      \label{fig:ccv_vs_params}
    \end{figure*}

    \begin{itemize}

      \item Lens redshift (\(z_{\rm l}\)) affects the lens-source distance ratio \(D_{ls}/D_s\) and thus the redshift range over which HU configurations can form \citep{schneider1992}.  
      
      \item Einstein radius (\(R_E\)) and Ellipticity (\(e\)), as defined in Section~\ref{sec:sample}, play a central role in shaping caustic geometry and, by extension, the probability of HU formation. Specifically, the Einstein radius \(R_E\) sets the overall size of strong-lensing regions, with larger \(R_E\) uniformly enlarging the caustic footprint \citep{kormann1994} while the ellipticity \(e\) elongates and distorts the caustic network, with higher \(e\) producing more stretched and complex critical curves \citep{meneghetti2007, kassiola1993}.

      \item Number of potentials (\(N_{\rm pot}\)) is defined as the total number of PIEMD components (including the central halo and sub-halos) located within the Einstein radius. Focusing on the inner lens region ensures that we capture only the substructures that are likely to interact with the critical curve and affect the formation of exotic caustic features. A higher \(N_{\rm pot}\) implies a more complex mass distribution near the lens centre, which increases the chance of topological perturbations such as cusp exchanges \citep{meneghetti2007, xu2009}. This motivation is supported by simulations showing that adding off-centre mass clumps near the Einstein radius can trigger the emergence of hyperbolic umbilic morphologies \citep{meena2023exotic}.

      \item Substructure mass fraction (\(f_{\rm sub}\)) quantifies the proportion of the cluster's lensing mass arising from galaxy-scale sub-halos. In each \texttt{Lenstool} model, we identify sub-halos as PIEMD potentials with velocity dispersion \(\sigma < 500\ \mathrm{km\,s^{-1}}\) and truncation radius \(r_{\rm cut} < 200\ \mathrm{kpc}\), to cover the typical range of galaxy potentials generally found in previous studies \citep{Richard2021, fox2022strongest}. We generate two convergence-based mass maps: one including all potentials (\(M_{\rm all}\)), and one excluding these sub-halos (\(M_{\rm smooth}\)). 

      To ensure homogeneity across clusters at different redshifts, we integrate both masses within a fixed physical aperture of 500~kpc centred on the cluster. This consistent aperture avoids redshift-dependent biases and focuses on the inner region where strong-lensing occurs. We then define the substructure mass fraction as
      
      \begin{equation}
        f_{\rm sub} \;=\;\frac{M_{\rm all} - M_{\rm smooth}}{M_{\rm all}}.
        \label{eq:fsub}
      \end{equation}
      
      This quantity reflects the relative importance of small-scale mass perturbations. Larger values of \(f_{\rm sub}\) indicate a greater concentration of sub-halo mass, which can locally distort critical curves and increase the probability of exotic configurations such as hyperbolic umbilics \citep{meneghetti2007, mao1998}. At the same time, a higher \(f_{\rm sub}\) also implies a more complex lensing potential, making HU image reconstruction and model optimisation more challenging, since a larger number of small halos must be explicitly constrained in the mass model.

      \item Cuspiness (\(\Delta\eta\)) measures the change in steepness of the convergence field. Based on maps evaluated at \(D_{ls}/D_s = 1\), ensuring a consistent normalisation of \(\kappa\) across clusters, we first define a set of ten logarithmically spaced target radii \(r_{\rm targ}\) within two regions: the inner core (\(0.05\,R_{\rm Ein} < r_{\rm targ} < 0.1\,R_{\rm Ein}\)) and the outer region (\(R_{\rm Ein} < r_{\rm targ} < 1.5\,R_{\rm Ein}\)). For each \(r_{\rm targ}\), a dichotomy search finds the level \(\kappa_0\) whose closed \(\kappa\)-isocontour yields this specific radius value from its enclosed area, and we then compute the mean convergence \(\bar\kappa\) inside. This approach captures the true 2D morphology, allowing halo ellipticity and substructure to shape the structure of the cluster, rather than assuming perfect circular symmetry \citep{umetsu2016}.
      
      A robust RANSAC (Random Sample Consensus) fit of 
      \(\ln\bar\kappa = a\,\ln r + b\) 
      is performed to estimate the logarithmic slope of the convergence profile within each region. RANSAC iteratively fits a model to random subsets of the data and retains the solution that maximises the number of inliers, thereby reducing the influence of outliers. Ultimately, we define the cuspiness of the cluster as

      \begin{equation}
        \Delta\eta = \eta_{\rm in} - \eta_{\rm out}
        \quad \text{with} \quad
        \eta = -\frac{\mathrm{d}\ln\kappa}{\mathrm{d}\ln r}.
      \label{eq:cuspiness}
      \end{equation}

      Uncertainties \(\sigma_{\Delta\eta}\) are obtained through error propagation from the RANSAC fit covariance \citep{fischler1981}. Since the central concentration and relaxation state are known to affect the geometry of critical curves \citep{newman2013}, we expect cuspiness to significantly influence the emergence and geometry of HU configurations.
      
    \end{itemize}
    
    \begin{figure*}[!ht]
      \centering
      \includegraphics[width=\textwidth]{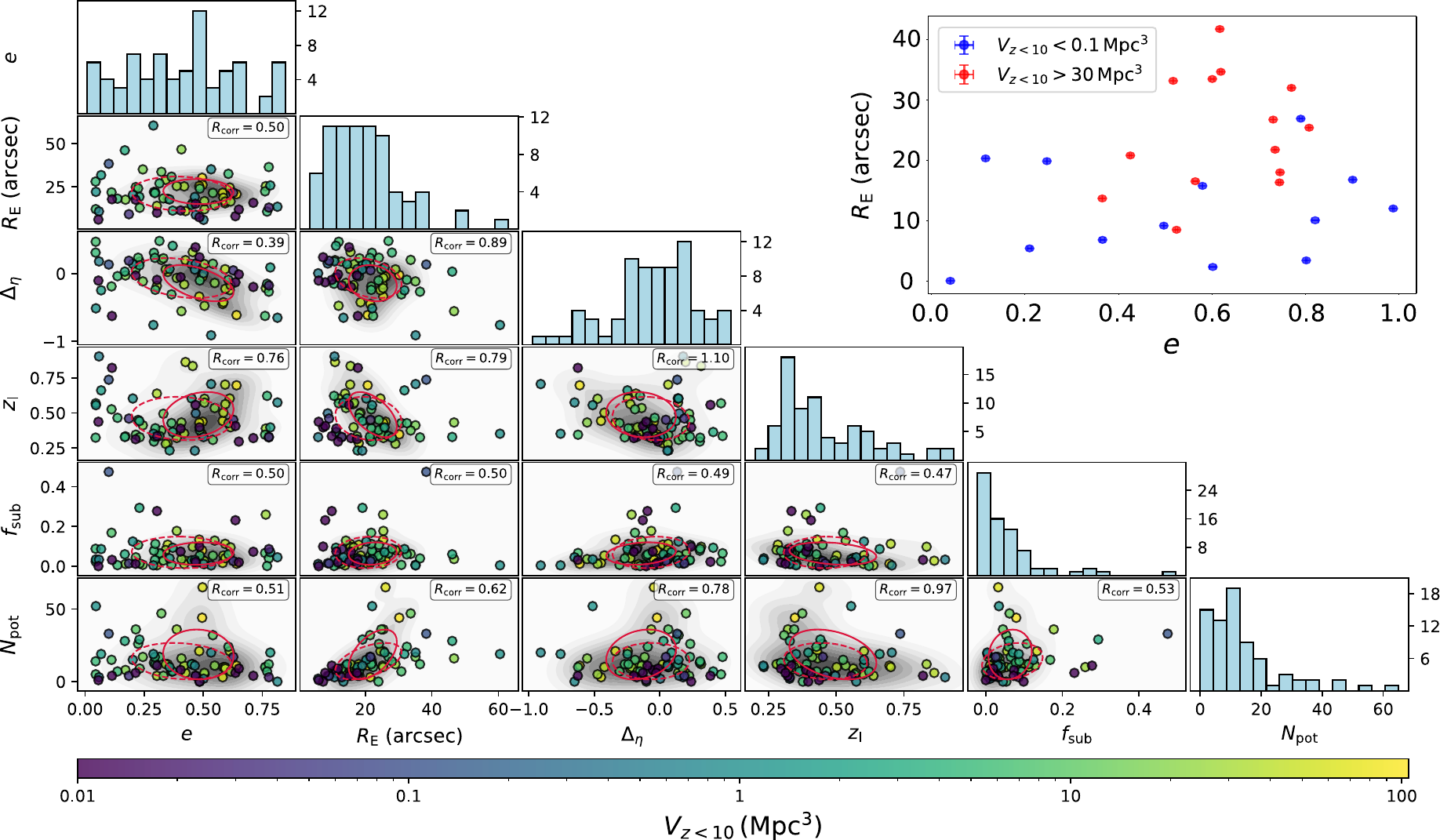}
      \caption{Triangle plot of $V_{z<10}$ against all parameter pairs. Off-diagonal panels show $V_{z<10}$-coloured scatter, weighted KDE, 1-$\sigma$ ellipses (dashed: unweighted, solid: $V_{z<10}$-weighted), and the associated correlation ratio $R_{\rm corr}$; diagonal panels show histograms for each parameter. The top-right panel illustrates the distribution of clusters in the $(e, R_{\rm E})$ plane, with blue and red points highlighting two extreme populations with lowest and highest $V_{z<10}$ values, respectively.
}
      \label{fig:triangle_ccv}
    \end{figure*}

    To probe whether any of these parameters is directly related to the total exotic comoving volume \(V_{z<10}\), we have produced a six-panel figure (Figure~\ref{fig:ccv_vs_params}) in which \(V_{z<10}\) is plotted against each of them. In each panel, points are shown with horizontal error bars whenever an uncertainty is available (\(R_E\), \(\sigma_e\), and \(\sigma_{\Delta\gamma}\)). Points without error bars denote parameters directly estimated from the Lenstool models and produced maps. A quantile binning into \(N_{\rm bins}=10\) equal-occupancy bins along the horizontal axis allows us to compute a typical mean value \(\langle \mathrm{V_{z<10}}\rangle\) in each interval, with standard errors on the mean \(\mathrm{SE}(\mathrm{V_{z<10}}) = \sigma_{\mathrm{V_{z<10}}} / \sqrt{N_{\rm bin}}\).

    Figure~\ref{fig:ccv_vs_params} shows that none of the six structural parameters considered in isolation displays a strong correlation with the exotic volume. However, among them, ellipticity stands out as the most informative: clusters with moderate to high ellipticities (\(e \sim 0.7\)) tend to produce systematically larger exotic regions. This trend, though not strictly monotonic, is clearly discernible and suggests a physical link. It is also consistent with the findings of \citet{meena2024image}, who show that elongated mass distributions favour the formation of stable hyperbolic umbilic configurations, with enhanced separation and visibility of the central images.

    It is important to remember that, even though our cluster sample appears morphologically diverse, as shown in Figure~\ref{histo_morpho}, it is not uniformly distributed in all parameters. In particular, systems with very large Einstein radii are slightly underrepresented, which could limit our ability to detect clearer trends. This is especially relevant given our expectations that larger \(R_{\rm E}\) values should enhance caustic extent and thus the likelihood of HU exotic image formation.

    Overall, this first exploration of the structural parameter space provides a meaningful and physically motivated framework for looking for conditions under which HU exotic lenses emerge, also suggesting that HU formation depends on a more intricate interplay of cluster properties, motivating an exploration of two-dimensional parameter spaces.

    Figure~\ref{fig:triangle_ccv} presents all pairwise combinations among the six parameters \( \{e,\,R_E,\,\Delta\eta,\,z_{\rm l},\,f_{\rm sub},\,N_{\rm pot}\} \) against the exotic comoving volume \(V_{z<10}\). In each off-diagonal panel, clusters are plotted as points, coloured by \(V_{z<10}\), overlaying a greyscale 2D Gaussian kernel density estimate weighted by \(V_{z<10}\), computed with $gaussian\_kde$ (from SciPy), to highlight where the highest-volume systems concentrate \citep{silverman2018}.

    To quantify how well each pair of parameters isolates the most exotic clusters, we compute two \(2\times2\) covariance matrices in each two-parameter space: one unweighted and one weighted by \(w_k = \mathrm{V_{z<10,\,k}} / \sum_k \mathrm{V_{z<10,\,k}}\) \citep{mardia2024}. We then overlay their corresponding 1-\(\sigma\) ellipses (dashed for the unweighted sample, solid for the \(V_{z<10}\)-weighted sample). Extracting eigenvalues \(\lambda_{1,\mathrm{uwt}}\ge\lambda_{2,\mathrm{uwt}}\) from the unweighted covariance matrix and \(\lambda_{1,\mathrm{wt}}\ge\lambda_{2,\mathrm{wt}}\) from the weighted covariance matrix gives the ellipse area scales as \(\pi\sqrt{\lambda_1\lambda_2}\) and the minor-to-major axis ratio as \(\sqrt{\lambda_2/\lambda_1}\). We then define the correlation ratio

    \begin{equation}
      R_{\rm corr}
      =
      \frac{\sqrt{\lambda_{1,\mathrm{wt}}\,\lambda_{2,\mathrm{wt}}}}
          {\sqrt{\lambda_{1,\mathrm{uwt}}\,\lambda_{2,\mathrm{uwt}}}}
      \;\times\;
      \frac{\sqrt{\lambda_{2,\mathrm{wt}}/\lambda_{1,\mathrm{wt}}}}
          {\sqrt{\lambda_{2,\mathrm{uwt}}/\lambda_{1,\mathrm{uwt}}}},
    \label{eq:corr_ratio}
    \end{equation}

    which falls well below unity only when the \(V_{z<10}\)-weighted ellipse is both substantially smaller in area and/or more elongated than the unweighted ellipse, i.e.\ when that parameter pair best distinguishes the highest-exotic volume clusters.

    We find the smallest correlation ratios, and hence the strongest discriminating power for high-\(V_{z<10}\) systems, in the following parameter pairs:
    \[
    R_{\rm corr}(e,\Delta_{\eta}) = 0.39 \\
    R_{\rm corr}(z_l,f_{\rm sub}) = 0.47 \\
    R_{\rm corr}(\Delta_{\eta},f_{\rm sub}) = 0.49
    \]

    Overall, we find that parameter pairs involving either ellipticity or the substructure mass fraction within \(R_E\) tend to yield relatively low correlation ratios \(R_{\rm corr}\), indicating better discriminative power. Among these, the pair \((\Delta\eta, e)\) emerges as the most effective indicator: the most exotic clusters are typically characterised by moderate ellipticities (\(e \sim 0.5\)) and no cuspiness (\(\Delta\eta \approx 0\)). This suggests that such structural characteristics are particularly favourable for generating extended exotic lensing regions.

    Beyond this best-performing pair, other combinations also provide useful discrimination: \((e, R_E)\), with a correlation ratio of 0.5, but also \((e, f_{\rm sub})\) and \((R_E, f_{\rm sub})\). This trend is illustrated in the top-right panel of Figure~\ref{fig:triangle_ccv}, where we contrast the clusters with the highest and lowest \(V_{z<10}\) values from our sample, each group containing the same number of clusters, in the \((e, R_E)\) plane. The two populations appear distinct, reinforcing the idea that even moderate ellipticity combined with a large Einstein radius is not only a useful predictor but in practice almost a requirement for efficient exotic lensing. Indeed, very high ellipticity values tend to suppress the exotic volume \(V_z\) across our sample, consistent with the trend reported by \citet{meena2023exotic}. This also confirms that parameter pairs with \(R_{\rm corr} \leq 0.5\) retain significant power to distinguish between exotic and non-exotic clusters.

    We also investigated whether similar separability arises when contrasting clusters based on the shape of their exotic area profile, distinguishing between systems with sharply peaked \(\Omega(z)\) and those exhibiting broader, more monotonic growth (Figure~\ref{surface_vs_z}). However, no clear distinction emerges in the bivariate parameter spaces considered. Finally, we note that repeating the full analysis with \(V_{z<4}\) as the weighting factor yields consistent patterns, underscoring the robustness of these conclusions across source redshift intervals. A natural extension would be to explore the full multivariate parameter space, for instance through a principal component analysis (PCA), to identify the most discriminating parameter combinations; we defer this investigation to future work.

\subsection{Expected number of HU exotic configurations in each cluster}\label{sec:HU_number}

    Now that we have estimated the exotic comoving volume associated with each cluster, we can turn to predicting the minimal number of galaxies expected to lie within these regions. This exercise provides a concrete sense of what observations might realistically reveal, given standard cosmological assumptions and measured galaxy populations.

    As a consistency check on our exotic volume predictions, we compare our results to independent estimates of the number of observable HU configurations available in the literature. Early ray-tracing work by \citet{Xivry2009} suggests an abundance of roughly one HU configuration per full-sky survey of cluster lenses, implying a very low surface density on the sky. More recently, \citet{meena2021} estimated that deep JWST observations should reveal at least one HU and one swallowtail configuration for every five high-redshift clusters.

    Several exotic HU systems have already been observationally confirmed, such as RXJ0437+00 and A1703, which we have analysed in earlier sections. Beyond these, additional candidates have been reported in various observational studies. Some of these remain qualitative in nature. For instance, \citet{Richard2021} note that several clusters in their sample exhibit lensing geometries potentially consistent with HU lensing, suggesting that such features may be more widespread than the currently limited number of confirmed cases implies. Similarly, \citet{ebeling2025} highlight eMACSJ1248.2 as displaying a HU-like configuration, based on visual inspection. More concrete candidates include a HU-like configuration which has been identified in the $z = 0.3$ Herschel field studied by \citet{egami2012}, where a submillimeter galaxy at $z = 4.69$ appears resolved into five images consistent with hyperbolic umbilic morphology. Likewise, \citet{tagliavia2024} reported a candidate HU system in COOL J0745+0924, a $z = 3.55$ galaxy lensed by a $z \sim 1$ cluster.

    Beyond these isolated detections and mentions, more systematic identifications of promising clusters have recently emerged. \citet{meena2022exotic} presented a list of eight clusters from the RELICS survey that exhibit large HU cross-sections, based on detailed ray-tracing analyses. Notably, three of these (RXJ0437, MACS0257, and MACS0417) are part of our current sample and were independently recognised in our analysis as particularly favourable environments for exotic HU formation. Further support comes from \citet{meena2023exotic}, who report three new potential HU systems in Abell 2163 and SPT0615.

    Altogether, these findings underscore that the expected number of HU lenses remains uncertain, reinforcing the need for systematic, quantitative approaches. The method developed in this work directly addresses this need by providing a framework to estimate their statistical occurrence.

    To first order, the expected number of HU configurations scales directly with the exotic comoving volume, since sampling a larger volume simply increases the number of background sources and thus the probability of HU formation. However, we can have cluster-to-cluster differences tied to survey depth because both the exotic area and the background galaxy density, set by the evolving luminosity function \citep{finkelstein2015}, vary strongly with redshift. 
    
    In particular, clusters whose exotic regions lie predominantly at low redshift probe smaller comoving volumes per unit solid angle \citep{hogg1999}, whereas high-redshift exotic regions benefit from larger volumes but require sufficiently deep imaging to reveal faint background galaxies.

    To proceed, we compute galaxy counts over the redshift range where the ultraviolet luminosity function (UV LF) is well-constrained by observations. Specifically, we adopt redshift bins from \( z = 1 \) to \( z = 4 \), excluding lower redshifts where rest-frame UV is not reliably probed in most datasets. Each bin is centred at a mean redshift \( \bar{z} \), and we extract the corresponding exotic comoving volume \( \mathrm{d}V(\bar{z}) \) for use in our calculation.

    Within each bin, we assume that galaxies follow a Schechter ultraviolet luminosity function, parameterised by a redshift-dependent characteristic magnitude \( M^*(\bar{z}) \), normalisation \( \phi^*(\bar{z}) \), and faint-end slope \( \alpha(\bar{z}) \) \citep{schechter1976}.

    These parameters are adopted from the fiducial parametric model of \citealp{bouwens2021}, based on both HFF and blank-field surveys. The luminosity function is taken to be constant within each bin; to justify this, we ensured that our binning in $D_{ls}/D_s$ is fine enough to minimise systematic errors from redshift evolution across individual bins. The number density of galaxies per unit magnitude at redshift \(\bar{z}\) is given by

    \begin{multline}
      \frac{\mathrm{d}n}{\mathrm{d}m}(m,\bar{z})
      = (0.4 \ln 10)\,\phi^*(\bar{z})\;\times\\
      10^{0.4\bigl(M^*(\bar{z}) - M(m,\bar{z})\bigr)\,(\alpha(\bar{z}) + 1)}
      \,\exp\!\Bigl[-10^{0.4\bigl(M^*(\bar{z}) - M(m,\bar{z})\bigr)}\Bigr],
    \label{eq:number_density}
    \end{multline}
    where \(M(m,\bar{z}) = m - \mathrm{d_m}(\bar{z})\) is the absolute magnitude corresponding to apparent magnitude \(m\), and \(\mathrm{d_m}(\bar{z})\) is the distance modulus.  
    
    We can obtain the number density of galaxies brighter than a limiting apparent AB magnitude \(m_{\lim}\) at redshift \(\bar{z}\) by integrating the Schechter luminosity function
    \begin{equation}
      \frac{\mathrm{d}N}{\mathrm{d}V_{\rm com}}\bigl(<m_{\lim},\bar{z}\bigr)
      = \int_{-\infty}^{m_{\lim}} \frac{\mathrm{d}n}{\mathrm{d}m}(m,\bar{z})\,\mathrm{d}m.
      \label{eq:number_density}
    \end{equation}

    Multiplying this number density by the exotic comoving volume element and integrating over redshift then yields the total expected number of galaxies within the exotic volume of a given cluster:
    \begin{equation}
      N_{\rm gal}(<m_{\lim}) = \int_{z=1}^{4} 
      \left[
      \frac{\mathrm{d}N}{\mathrm{d}V_{\rm com}}(<m_{\lim},z)\; \frac{\mathrm{d}V_{\rm com}}{\mathrm{d}z}
      \right]\,\mathrm{d}z.
      \label{eq:ngal_integral}
    \end{equation}

    The limiting magnitude \(m_{\mathrm{lim}}\) used for each cluster reflects the observational depth of the corresponding dataset. Since the clusters come from different surveys, this limiting magnitude varies accordingly. For instance, a depth of \(m_{\mathrm{lim}} \approx 27.5\) is typical for CLASH clusters or HST snapshots, whereas Hubble Frontier Fields can reach up to \(m_{\mathrm{lim}} \approx 29\). These values are specified for each cluster in Table~\ref{tab:complete}.

    Importantly, this method does not account for the increased brightness due to lensing magnification. Since we aim to compare the expected numbers to actual observed HU systems, including their less magnified counter-images, this ensures our estimates remain conservative and represent strict lower limits.

\begin{figure}[ht]
    \centering
    \includegraphics[width=\columnwidth]{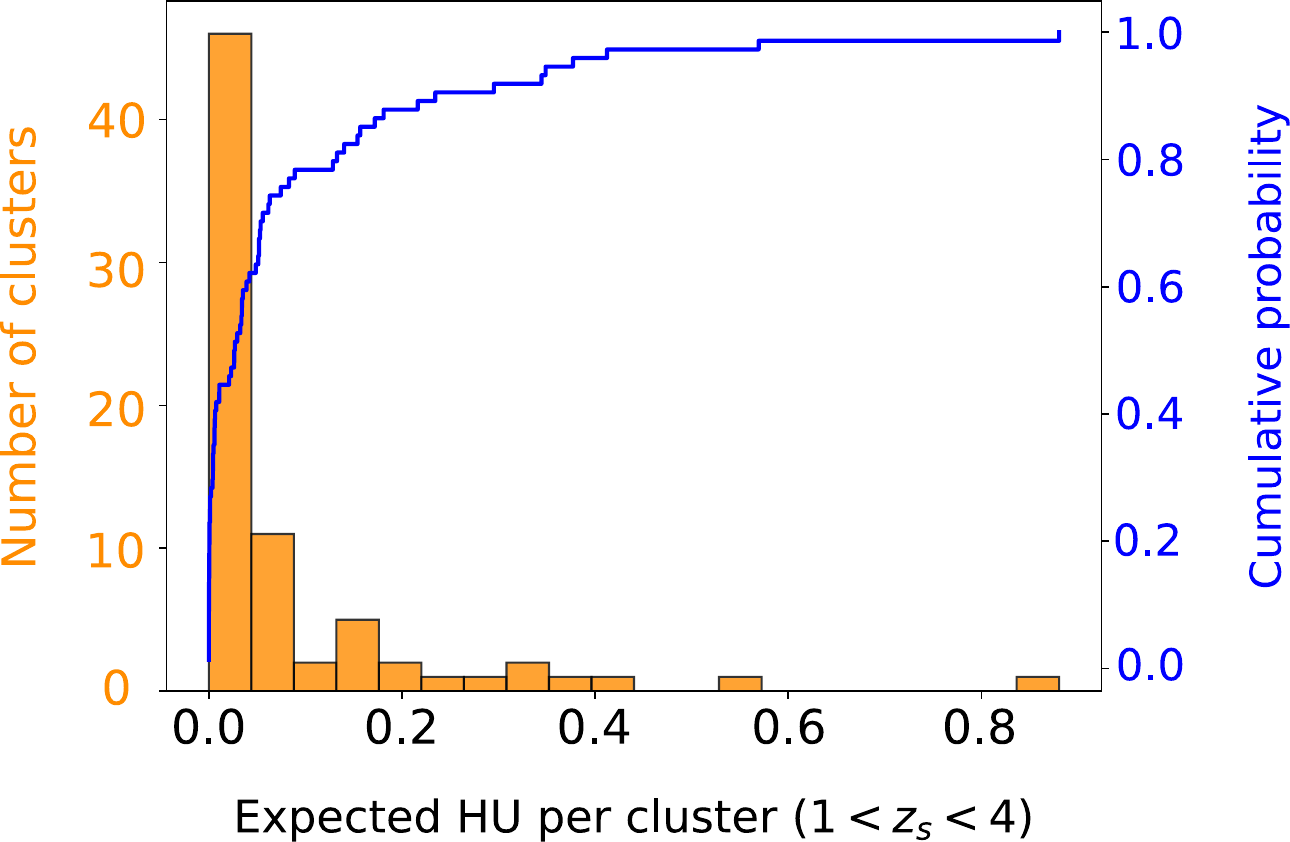}
    \caption{Distribution (orange) and empirical cumulative distribution function (CDF, blue) of the expected number of observable HU systems per cluster, \(N_{\rm gal}(<m_{\lim})\), computed across the 74-cluster sample and considering a source redshift $1 < z_s < 4$.}
    \label{fig:CDF}
    \end{figure}

    Across our sample of 74 clusters, the per-cluster expected counts \(N_{\rm gal}(<m_{\lim})\) range from 0 (the least favourable fields) up to \(\sim0.9\) observable HU systems, with a median of 0.027, a mean of \(\langle N_{\rm gal} \rangle = 0.081\), and a standard deviation \(\sigma = 0.147\). Figure~\ref{fig:CDF} presents the histogram of \(N_{\rm gal}\), illustrating that most clusters contribute fewer than 0.1 galaxies within their exotic volumes, while a few high-yield clusters increase the mean.

    To examine the likelihood that a randomly selected cluster hosts at least a given number \(n\) of HU systems, Figure~\ref{fig:CDF} shows the empirical cumulative distribution function (CDF) of \(N_{\rm gal}\). While small features appear around \(N_{\rm gal}\approx0.1\), they are not statistically significant given the Poisson uncertainties. The overall trend simply indicates that only a minority of clusters host more than a handful of HU systems, underscoring the rarity of higher-yield fields.

    Assuming a Poisson process with mean rate \(\langle N_{\rm gal}\rangle\) per cluster, the probability of finding at least one exotic HU system among \(N\) independent clusters is
    \[
      P(\ge1 \mid N) \;=\; 1 - \exp\bigl(-N\,\langle N_{\rm gal}\rangle\bigr).
    \]
    Solving for \(N\) as a function of the desired detection probability \(P\) gives

    \begin{equation}
      N(P) \;=\; -\frac{\ln\bigl(1 - P\bigr)}{\langle N_{\rm gal}\rangle}.
      \label{eq:number_density}
    \end{equation}

    From this relation, one finds that approximately \(N=9\) clusters are required to achieve a 50\% chance of detecting at least one HU system, and \(N=29\) for 90\% confidence. These numbers naturally depend on the redshift depth considered for the source. For instance, extending the integration up to \(z < 10\) increases the expected number of HU systems to \(\sim 8.28\), thereby reducing the required number of clusters to \(N = 6\) and \(N = 20\), respectively, for 50\% and 90\% detection probabilities. Nevertheless, these results should be interpreted with some caution, as they are derived under the assumption of a statistically homogeneous sample, whereas our cluster set is intrinsically heterogeneous. The inferred probabilities should thus be viewed in the context of this diversity.

    Finally, summing \(N_{\rm gal}\) across all 74 selected clusters yields a total expected number of HU-candidate galaxies of \(\sum_{i=1}^{M} N_{\rm gal}^{(i)} \approx 6\), not accounting for lensing magnification. This theoretical lower bound already exceeds the four HU systems confirmed to date (one in A1703 and three in RXJ0437+00), and becomes even more consistent when including additional strong candidates from the literature, such as those proposed in \citet{meena2022exotic}, \citet{meena2023exotic}, and other observational works discussed above.
    
    Two main observational factors likely contribute to the apparent gap between expected and confirmed HU systems:
    
    \begin{enumerate}[label=\arabic*.]
      \item Lack of systematic HU searches. To date, no dedicated survey or analysis pipeline has been specifically designed to identify HU configurations. The few confirmed HU systems (as in Abell 1703; \citealp{Richard2009}) were uncovered serendipitously in studies optimised for classical strong-lensing features. Moreover, standard lens-modelling tools such as Lenstool \citep{jullo2007} are not tuned to detect these exotic configurations, suggesting that many HU systems may have gone unnoticed in existing data. Ongoing efforts aim to address this gap, including a systematic census of HU configurations currently in preparation (Lagattuta et al. in prep). 
      
      \item Incomplete multiwavelength coverage. Exotic HU sources are often revealing their strongest signals in specific bands (for instance narrowband Lyman-\(\alpha\)) and could be missed in the broadband optical/near-IR data currently available for many clusters (system~10 of RXJ0437+00 provides a striking example; \citealt{Lagattuta2023}). Deep integral-field spectroscopy, such as MUSE observations, has proven crucial for uncovering faint Lyman-\(\alpha\) emitters behind clusters \citep{bacon2021}. However, only a subset of our fields benefits from such data, underscoring the need for coordinated narrowband programs to fully characterise the HU exotic population.

    \end{enumerate}

    These observational limitations and selection effects suggest that the current sample of four confirmed HU systems likely represents a strict lower bound. As deep, multi-band and spectroscopic surveys increasingly target cluster fields, the number of detected exotic configurations is expected to rise towards the theoretical baseline.

    Ultimately, these results provide a clear and quantitative roadmap for designing cluster surveys aimed at uncovering HU exotic systems. A striking example is the GLIMPSE ultra-deep JWST NIRCam program on AS1063 (Program 3293; \citealp{kokorev2024}), which yields the deepest lensing fields to date and corresponds to one of the highest observed exotic volumes. Despite generally low per-cluster yields, focusing on a few dozen similarly deep fields, as demonstrated by GLIMPSE, should yield a high likelihood of detecting at least one HU exotic configuration, even before accounting for the amplifying effects of lensing magnification.

\section{Conclusions}
\label{sec:conclusion}
    In this work, we have addressed the dual challenge of the ambiguous definition and rare detection of HU exotic gravitational lenses. These systems, known for their unusual image configurations, strong magnifications, and near-isotropic distortions, hold great potential for studying dark-matter distributions and reconstructing distant sources. Yet their scarcity and the lack of a consensus definition have limited their use and interpretation in the literature.
    
    To move past these limitations, we developed a robust and physically motivated method to identify the galaxy clusters most likely to produce such HU exotic configurations. Relying on parametric mass models, our approach maps the regions in the source plane where a background object would be imaged exotically. By integrating these exotic areas across redshift, we define a cumulative comoving volume: a single, interpretable quantity that reflects the exotic predisposition of each cluster and enables direct comparison between them.

    We have tested and validated the robustness of our method against both modelling uncertainties and sampling stochasticity, ensuring that the derived exotic volumes are stable and reliable.

    Focusing on six structural parameters of clusters (\(e,\,R_E,\,\Delta\eta,\,z_{\rm l},\,f_{\rm sub},\,N_{\rm pot}\)), we found that none of them exhibits a clear one-dimensional correlation with the exotic comoving volume \(V_{z<10}\), suggesting that HU formation depends on a complex combination of cluster properties.
    We exhibited that two-parameter combinations, particularly ellipticity paired with Einstein radius, cuspiness, or number of potentials, emerge as the most effective discriminants for identifying the most exotic clusters.  

    By combining our computed exotic volumes with galaxy number densities from the Schechter luminosity function, while deliberately ignoring any lensing magnification to remain conservative, we find that each cluster hosts on average \(\approx0.125\) galaxies in its exotic region (with a maximum of \(\sim0.8\)). Under a Poisson model, achieving a 90\% probability of detecting at least one HU system therefore requires observations of roughly 19 such clusters. This conservative baseline provides a clear roadmap for future surveys: targeting on the order of ten to a few tens of deep cluster fields should secure a handful of HU exotic detections.  

    Ultimately, we propose the exotic comoving volume as a clear and physically meaningful metric to compare galaxy clusters and guide follow-up efforts. Looking ahead, further improvements in mass modelling, the inclusion of magnification effects, and the expansion to even larger cluster samples will contribute to building a more complete inventory of HU exotic gravitational lenses.

   \bibliographystyle{aa} 
   \bibliography{references}

@article{fox2022strongest,
  title={The strongest cluster lenses: an analysis of the relation between strong gravitational lensing strength and the physical properties of galaxy clusters},
  author={Fox, Carter and Mahler, Guillaume and Sharon, Keren and Gonz{\'a}lez, Juan D Remolina},
  journal={ApJ},
  volume={928},
  number={1},
  pages={87},
  year={2022},
  publisher={IOP Publishing}
}

@ARTICLE{Richard2021,
  title     = "An atlas of {MUSE} observations towards twelve massive lensing
               clusters",
  author    = "Richard, Johan and Claeyssens, Ad{\'e}la{\"\i}de and Lagattuta,
               David and Guaita, Lucia and Bauer, Franz Erik and Pello, Roser
               and Carton, David and Bacon, Roland and Soucail, Genevi{\`e}ve
               and Lyon, Gonzalo Prieto and Kneib, Jean-Paul and Mahler,
               Guillaume and Cl{\'e}ment, Benjamin and Mercier, Wilfried and
               Variu, Andrei and Tamone, Am{\'e}lie and Ebeling, Harald and
               Schmidt, Kasper B and Nanayakkara, Themiya and Maseda, Michael
               and Weilbacher, Peter M and Bouch{\'e}, Nicolas and Bouwens,
               Rychard J and Wisotzki, Lutz and de la Vieuville, Geoffroy and
               Martinez, Johany and Patr{\'\i}cio, Vera",
  journal   = "A\&A",
  publisher = "EDP Sciences",
  volume    =  646,
  pages     = "A83",
  month     =  feb,
  year      =  2021,
  copyright = "http://creativecommons.org/licenses/by/4.0"
}

@article{Xivry2009,
  title     = "An atlas of predicted exotic gravitational lenses",
  author    = "Orban de Xivry, Gilles and Marshall, Phil",
  journal   = "MNRAS",
  publisher = "Oxford University Press (OUP)",
  volume    =  399,
  number    =  1,
  pages     = "2--20",
  month     =  oct,
  year      =  2009,
  language  = "en"
}

@article{meena2020,
  title={Finding singularities in gravitational lensing},
  author={Meena, Ashish Kumar and Bagla, JS},
  journal={MNRAS},
  volume={492},
  number={3},
  pages={3294--3305},
  year={2020},
  publisher={Oxford University Press}
}

@book{schneider1992,
  title={Gravitational lenses as astrophysical tools},
  author={Schneider, Peter and Ehlers, J{\"u}rgen and Falco, Emilio E and Schneider, Peter and Ehlers, J{\"u}rgen and Falco, Emilio E},
  year={1992},
  publisher={Springer}
}

@article{bartelmann2001weak,
  title={Weak gravitational lensing},
  author={Bartelmann, Matthias and Schneider, Peter},
  journal={Physics Reports},
  volume={340},
  number={4-5},
  pages={291--472},
  year={2001},
  publisher={Elsevier}
}

@article{limousin2008,
  title={Strong lensing in Abell 1703: constraints on the slope of the inner dark matter distribution},
  author={Limousin, Marceau and Richard, Johan and Kneib, J-P and Brink, Henrik and Pell{\'o}, Roser and Jullo, Eric and Tu, Hong and Sommer-Larsen, Jesper and Egami, E and Micha{\l}owski, Micha{\l} J and others},
  journal={A\&A},
  volume={489},
  number={1},
  pages={23--35},
  year={2008},
  publisher={EDP Sciences}
}

@article{meena2021,
  title={Exotic image formation in strong gravitational lensing by clusters of galaxies--I. Cross-section},
  author={Meena, Ashish Kumar and Bagla, Jasjeet Singh},
  journal={MNRAS},
  volume={503},
  number={2},
  pages={2097--2107},
  year={2021},
  publisher={Oxford University Press}
}

@article{jullo2007,
  title={A Bayesian approach to strong lensing modelling of galaxy clusters},
  author={Jullo, Eric and Kneib, Jean-Paul and Limousin, Marceau and Eliasdottir, Ardis and Marshall, PJ and Verdugo, Tomas},
  journal={New Journal of Physics},
  volume={9},
  number={12},
  pages={447},
  year={2007},
  publisher={IOP Publishing}
}

@article{natarajan1997,
  title={Lensing by galaxy haloes in clusters of galaxies},
  author={Natarajan, Priyamvada and Kneib, Jean-Paul},
  journal={MNRAS},
  volume={287},
  number={4},
  pages={833--847},
  year={1997},
  publisher={Blackwell Science Ltd Oxford, UK}
}

@ARTICLE{Lotz2017,
       author = {{Lotz}, J.~M. and {Koekemoer}, A. and {Coe}, D. and {Grogin}, N. and {Capak}, P. and {Mack}, J. and {Anderson}, J. and {Avila}, R. and {Barker}, E.~A. and {Borncamp}, D. and {Brammer}, G. and {Durbin}, M. and {Gunning}, H. and {Hilbert}, B. and {Jenkner}, H. and {Khandrika}, H. and {Levay}, Z. and {Lucas}, R.~A. and {MacKenty}, J. and {Ogaz}, S. and {Porterfield}, B. and {Reid}, N. and {Robberto}, M. and {Royle}, P. and {Smith}, L.~J. and {Storrie-Lombardi}, L.~J. and {Sunnquist}, B. and {Surace}, J. and {Taylor}, D.~C. and {Williams}, R. and {Bullock}, J. and {Dickinson}, M. and {Finkelstein}, S. and {Natarajan}, P. and {Richard}, J. and {Robertson}, B. and {Tumlinson}, J. and {Zitrin}, A. and {Flanagan}, K. and {Sembach}, K. and {Soifer}, B.~T. and {Mountain}, M.},
        title = "{The Frontier Fields: Survey Design and Initial Results}",
      journal = {\apj},
         year = 2017,
        month = mar,
       volume = {837},
       number = {1},
          eid = {97},
        pages = {97},
          doi = {10.3847/1538-4357/837/1/97},
archivePrefix = {arXiv},
       eprint = {1605.06567},
 primaryClass = {astro-ph.GA},
       adsurl = {https://ui.adsabs.harvard.edu/abs/2017ApJ...837...97L}
}

@ARTICLE{Richard2009,
       author = {{Richard}, J. and {Pei}, L. and {Limousin}, M. and {Jullo}, E. and {Kneib}, J.~P.},
        title = "{Keck spectroscopic survey of strongly lensed galaxies in Abell 1703: further evidence of a relaxed, unimodal cluster}",
      journal = {\aap},
         year = 2009,
        month = apr,
       volume = {498},
       number = {1},
        pages = {37-47},
          doi = {10.1051/0004-6361/200811366},
archivePrefix = {arXiv},
       eprint = {0901.0427},
 primaryClass = {astro-ph.GA},
       adsurl = {https://ui.adsabs.harvard.edu/abs/2009A&A...498...37R}
}

@article{zitrin2026strong,
  title={Strong Gravitational Lensing with the James Webb Space Telescope},
  author={Zitrin, Adi},
  journal={arXiv preprint arXiv:2605.15189},
  year={2026}
}

@article{bergamini2025euclid,
  title={Euclid Quick Data Release (Q1). The first catalogue of strong-lensing galaxy clusters},
  author={Bergamini, P and Meneghetti, M and Acebron, A and Cl{\'e}ment, B and Bolzonella, M and Grillo, C and Rosati, P and Abriola, D and Barroso, JA Acevedo and Angora, G and others},
  journal={A\&A},
  year={2025},
  publisher={EDP Sciences}
}

@article{Ebeling_2007,
doi = {10.1086/518603},
url = {https://dx.doi.org/10.1086/518603},
year = {2007},
month = {may},
publisher = {},
volume = {661},
number = {1},
pages = {L33},
author = {Ebeling, H. and Barrett, E. and Donovan, D. and Ma, C.-J. and Edge, A. C. and van Speybroeck, L.},
title = {A Complete Sample of 12 Very X-Ray Luminous Galaxy Clusters at z &gt; 0.5},
journal = {ApJ}
}

@article{repp2018,
  title={Science from a glimpse: Hubble SNAPshot observations of massive galaxy clusters},
  author={Repp, A and Ebeling, H},
  journal={MNRAS},
  volume={479},
  number={1},
  pages={844--864},
  year={2018},
  publisher={Oxford University Press}
}

@article{ebeling2025,
  title={Beyond MACS: physical properties of extremely X-ray luminous clusters at z> 0.5},
  author={Ebeling, H and Richard, J and Beauchesne, B and Basto, Q and Edge, A\_C and Smail, I},
  journal={MNRAS},
  volume={537},
  number={3},
  pages={2662--2694},
  year={2025},
  publisher={Oxford University Press}
}

@article{richard2010,
  title={LoCuSS: first results from strong-lensing analysis of 20 massive galaxy clusters at z= 0.2},
  author={Richard, Johan and Smith, Graham P and Kneib, Jean-Paul and Ellis, Richard S and Sanderson, Alastair JR and Pei, Liuyi and Targett, TA and Sand, DJ and Swinbank, AM and Dannerbauer, Helmut and others},
  journal={MNRAS},
  volume={404},
  number={1},
  pages={325--349},
  year={2010},
  publisher={Blackwell Publishing Ltd Oxford, UK}
}

@article{postman2012cluster,
  title={The cluster lensing and supernova survey with Hubble: an overview},
  author={Postman, Marc and Coe, Dan and Benitez, Narciso and Bradley, Larry and Broadhurst, Tom and Donahue, Megan and Ford, Holland and Graur, Or and Graves, Genevieve and Jouvel, Stephanie and others},
  journal={A\&AS},
  volume={199},
  number={2},
  pages={25},
  year={2012},
  publisher={The American Astronomical Society}
}

@article{coe2019,
  title={RELICS: reionization lensing cluster survey},
  author={Coe, Dan and Salmon, Brett and Brada{\v{c}}, Maru{\v{s}}a and Bradley, Larry D and Sharon, Keren and Zitrin, Adi and Acebron, Ana and Cerny, Catherine and Cibirka, Nath{\'a}lia and Strait, Victoria and others},
  journal={ApJ},
  volume={884},
  number={1},
  pages={85},
  year={2019},
  publisher={IOP Publishing}
}

@article{claeyssens2022,
  title={The Lensed Lyman-Alpha MUSE Arcs Sample (LLAMAS)-I. Characterisation of extended Lyman-alpha halos and spatial offsets},
  author={Claeyssens, Ad{\'e}la{\"\i}de and Richard, Johan and Blaizot, J{\'e}r{\'e}my and Garel, Thibault and Kusakabe, Haruka and Bacon, Roland and Bauer, Franz E and Guaita, Lucia and Jeanneau, Alexandre and Lagattuta, D and others},
  journal={A\&A},
  volume={666},
  pages={A78},
  year={2022},
  publisher={EDP Sciences}
}

@article{meneghetti2023,
  title={A persistent excess of galaxy-galaxy strong lensing observed in galaxy clusters},
  author={Meneghetti, Massimo and Cui, Weiguang and Rasia, Elena and Yepes, Gustavo and Acebron, Ana and Angora, Giuseppe and Bergamini, Pietro and Borgani, Stefano and Calura, Francesco and Despali, Giulia and others},
  journal={A\&A},
  volume={678},
  pages={L2},
  year={2023},
  publisher={EDP Sciences}
}

@article{richard2014,
  title={Mass and magnification maps for the Hubble Space Telescope Frontier Fields clusters: implications for high-redshift studies},
  author={Richard, Johan and Jauzac, Mathilde and Limousin, Marceau and Jullo, Eric and Cl{\'e}ment, Benjamin and Ebeling, Harald and Kneib, Jean-Paul and Atek, Hakim and Natarajan, Priya and Egami, Eiichi and others},
  journal={MNRAS},
  volume={444},
  number={1},
  pages={268--289},
  year={2014},
  publisher={The Royal Astronomical Society}
}

@article{kawamata2018,
  title={Size--luminosity relations and UV luminosity functions at z= 6--9 simultaneously derived from the complete Hubble Frontier Fields data},
  author={Kawamata, Ryota and Ishigaki, Masafumi and Shimasaku, Kazuhiro and Oguri, Masamune and Ouchi, Masami and Tanigawa, Shingo},
  journal={ApJ},
  volume={855},
  number={1},
  pages={4},
  year={2018},
  publisher={IOP Publishing}
}

@article{johnson2014,
  title={Lens models and magnification maps of the six Hubble Frontier Fields clusters},
  author={Johnson, Traci L and Sharon, Keren and Bayliss, Matthew B and Gladders, Michael D and Coe, Dan and Ebeling, Harald},
  journal={ApJ},
  volume={797},
  number={1},
  pages={48},
  year={2014},
  publisher={IOP Publishing}
}

@article{raney2020,
  title={Exploring effects on magnifications due to line-of-sight galaxies in the Hubble Frontier Fields},
  author={Raney, Catie A and Keeton, Charles R and Brennan, Sean},
  journal={MNRAS},
  volume={492},
  number={1},
  pages={503--527},
  year={2020},
  publisher={Oxford University Press}
}

@article{umetsu2016,
  title={CLASH: Joint analysis of strong-lensing, weak-lensing shear, and magnification data for 20 galaxy clusters},
  author={Umetsu, Keiichi and Zitrin, Adi and Gruen, Daniel and Merten, Julian and Donahue, Megan and Postman, Marc},
  journal={ApJ},
  volume={821},
  number={2},
  pages={116},
  year={2016},
  publisher={IOP Publishing}
}

@article{fischler1981,
  title={Random sample consensus: a paradigm for model fitting with applications to image analysis and automated cartography},
  author={Fischler, Martin A and Bolles, Robert C},
  journal={Communications of the ACM},
  volume={24},
  number={6},
  pages={381--395},
  year={1981},
  publisher={ACM New York, NY, USA}
}

@article{newman2013,
  title={The density profiles of massive, relaxed galaxy clusters. II. Separating luminous and dark matter in cluster cores},
  author={Newman, Andrew B and Treu, Tommaso and Ellis, Richard S and Sand, David J},
  journal={ApJ},
  volume={765},
  number={1},
  pages={25},
  year={2013},
  publisher={IOP Publishing}
}

@article{meneghetti2007,
  title={Arc sensitivity to cluster ellipticity, asymmetries, and substructures},
  author={Meneghetti, Massimo and Argazzi, Rodolfo and Pace, Francesco and Moscardini, Lauro and Dolag, Klaus and Bartelmann, Matthias and Li, Guoliang and Oguri, Masamune},
  journal={A\&A},
  volume={461},
  number={1},
  pages={25--38},
  year={2007},
  publisher={EDP Sciences}
}

@article{mao1998,
  title={Evidence for substructure in lens galaxies?},
  author={Mao, Shude and Schneider, Peter},
  journal={MNRAS},
  volume={295},
  number={3},
  pages={587--594},
  year={1998},
  publisher={The Royal Astronomical Society}
}

@article{kassiola1993,
  title={Elliptic mass distributions versus elliptic potentials in gravitational lenses},
  author={Kassiola, Aggeliki and Kovner, Israel},
  journal={ApJ},
  volume={417},
  pages={450},
  year={1993}
}

@article{kormann1994,
  title={Isothermal elliptical gravitational lens models},
  author={Kormann, Robert and Schneider, Peter and Bartelmann, Matthias},
  journal={A\&A},
  volume={284},
  pages={285--299},
  year={1994}
}

@article{xu2009,
  title={Effects of dark matter substructures on gravitational lensing: results from the Aquarius simulations},
  author={Xu, DD and Mao, Shude and Wang, Jie and Springel, V and Gao, Liang and White, SDM and Frenk, Carlos S and Jenkins, Adrian and Li, Guoliang and Navarro, Julio F},
  journal={MNRAS},
  volume={398},
  number={3},
  pages={1235--1253},
  year={2009},
  publisher={Blackwell Publishing Ltd Oxford, UK}
}

@book{silverman2018,
  title={Density estimation for statistics and data analysis},
  author={Silverman, Bernard W},
  year={2018},
  publisher={Routledge}
}

@book{mardia2024,
  title={Multivariate analysis},
  author={Mardia, Kanti V and Kent, John T and Taylor, Charles C},
  year={2024},
  publisher={John Wiley \& Sons}
}

@article{navarro1997,
  title={A universal density profile from hierarchical clustering},
  author={Navarro, Julio F and Frenk, Carlos S and White, Simon DM},
  journal={ApJ},
  volume={490},
  number={2},
  pages={493},
  year={1997},
  publisher={IOP Publishing}
}

@article{eliasdottir2007,
  title={Where is the matter in the Merging Cluster Abell 2218?},
  author={El{\'\i}asd{\'o}ttir, {\'A}rd{\'\i}s and Limousin, Marceau and Richard, Johan and Hjorth, Jens and Kneib, Jean-Paul and Natarajan, Priya and Pedersen, Kristian and Jullo, Eric and Paraficz, Danuta},
  journal={arXiv preprint arXiv:0710.5636},
  year={2007}
}

@article{finkelstein2015,
  title={The evolution of the galaxy rest-frame ultraviolet luminosity function over the first two billion years},
  author={Finkelstein, Steven L and Ryan, Russell E and Papovich, Casey and Dickinson, Mark and Song, Mimi and Somerville, Rachel S and Ferguson, Henry C and Salmon, Brett and Giavalisco, Mauro and Koekemoer, Anton M and others},
  journal={ApJ},
  volume={810},
  number={1},
  pages={71},
  year={2015},
  publisher={IOP Publishing}
}

@article{hogg1999,
  title={Distance measures in cosmology},
  author={Hogg, David W},
  journal={arXiv preprint astro-ph/9905116},
  year={1999}
}

@article{bouwens2021,
  title={New determinations of the UV luminosity functions from z~ 9 to 2 show a remarkable consistency with halo growth and a constant star formation efficiency},
  author={Bouwens, RJ and Oesch, PA and Stefanon, M and Illingworth, G and Labb{\'e}, I and Reddy, N and Atek, H and Montes, M and Naidu, R and Nanayakkara, T and others},
  journal={ApJ},
  volume={162},
  number={2},
  pages={47},
  year={2021},
  publisher={IOP Publishing}
}

@article{schechter1976,
  title={An analytic expression for the luminosity function for galaxies.},
  author={Schechter, Paul},
  journal={ApJ},
  volume={203},
  pages={297--306},
  year={1976}
}

@article{bacon2021,
  title={The MUSE Extremely Deep Field: The cosmic web in emission at high redshift},
  author={Bacon, Roland and Mary, David and Garel, Thibault and Blaizot, Jeremy and Maseda, Michael and Schaye, Joop and Wisotzki, Lutz and Conseil, Simon and Brinchmann, Jarle and Leclercq, Floriane and others},
  journal={A\&A},
  volume={647},
  pages={A107},
  year={2021},
  publisher={EDP Sciences}
}

@article{kokorev2024,
  title={A Glimpse at the New Redshift Frontier Through Abell S1063},
  author={Kokorev, Vasily and Atek, Hakim and Chisholm, John and Endsley, Ryan and Chemerynska, Iryna and Mu{\~n}oz, Julian B and Furtak, Lukas J and Pan, Richard and Berg, Danielle and Fujimoto, Seiji and others},
  journal={arXiv preprint arXiv:2411.13640},
  year={2024}
}

@INPROCEEDINGS{tagliavia2024,
       author = {{Tagliavia}, Marie and {Horwath}, Gabriela and {Gladders}, Michael and {Cool-Lamps}},
        title = "{A Remarkable Hyperbolic Umbilic Strong Lens}",
    booktitle = {American Astronomical Society Meeting Abstracts \#243},
         year = 2024,
       series = {American Astronomical Society Meeting Abstracts},
       volume = {243},
        month = feb,
          eid = {460.02},
        pages = {460.02},
       adsurl = {https://ui.adsabs.harvard.edu/abs/2024AAS...24346002T}
}

@inproceedings{egami2012,
  title={Discovery of an Exceptionally Bright Gravitationally Lensed Submillimeter Galaxy at z= 4.69},
  author={Egami, Eiichi and others},
  booktitle={American Astronomical Society Meeting Abstracts\# 220},
  volume={220},
  pages={308--02},
  year={2012}
}

@article{meena2023exotic,
  title={Exotic image formation in strong gravitational lensing by clusters of galaxies--IV. Elliptical NFW lenses and hyperbolic umbilics},
  author={Meena, Ashish Kumar and Bagla, Jasjeet Singh},
  journal={MNRAS},
  volume={526},
  number={3},
  pages={3902--3919},
  year={2023},
  publisher={Oxford University Press}
}

@article{meena2021exotic,
  title={Exotic image formation in strong gravitational lensing by clusters of galaxies--II. Uncertainties},
  author={Meena, Ashish Kumar and Ghosh, Agniva and Bagla, Jasjeet S and Williams, Liliya LR},
  journal={MNRAS},
  volume={506},
  number={1},
  pages={1526--1539},
  year={2021},
  publisher={Oxford University Press}
}

@article{meena2022exotic,
  title={Exotic image formation in strong gravitational lensing by clusters of galaxies--III. Statistics with HUDF},
  author={Meena, Ashish Kumar and Bagla, Jasjeet Singh},
  journal={MNRAS},
  volume={515},
  number={3},
  pages={4151--4160},
  year={2022},
  publisher={Oxford University Press}
}

@article{meena2024image,
  title={Image formation near hyperbolic umbilic in strong gravitational lensing},
  author={Meena, Ashish Kumar and Bagla, Jasjeet Singh},
  journal={arXiv preprint arXiv:2405.16826},
  year={2024}
}

@ARTICLE{Lagattuta2023,
       author = {{Lagattuta}, David J. and {Richard}, Johan and {Ebeling}, Harald and {Basto}, Quentin and {Cerny}, Catherine and {Edge}, Alastair and {Jauzac}, Mathilde and {Mahler}, Guillaume and {Massey}, Richard},
        title = "{RXJ0437+00: constraining dark matter with exotic gravitational lenses}",
      journal = {\mnras},
         year = 2023,
        month = jun,
       volume = {522},
       number = {1},
        pages = {1091-1107},
          doi = {10.1093/mnras/stad803},
archivePrefix = {arXiv},
       eprint = {2303.09568},
 primaryClass = {astro-ph.CO},
       adsurl = {https://ui.adsabs.harvard.edu/abs/2023MNRAS.522.1091L}
}

@ARTICLE{Dessauges2017,
       author = {{Dessauges-Zavadsky}, M. and {Zamojski}, M. and {Rujopakarn}, W. and {Richard}, J. and {Sklias}, P. and {Schaerer}, D. and {Combes}, F. and {Ebeling}, H. and {Rawle}, T.~D. and {Egami}, E. and {Boone}, F. and {Cl{\'e}ment}, B. and {Kneib}, J. -P. and {Nyland}, K. and {Walth}, G.},
        title = "{Molecular gas properties of a lensed star-forming galaxy at z   3.6: a case study}",
      journal = {\aap},
         year = 2017,
        month = sep,
       volume = {605},
          eid = {A81},
        pages = {A81},
          doi = {10.1051/0004-6361/201628513},
archivePrefix = {arXiv},
       eprint = {1610.08065},
 primaryClass = {astro-ph.GA},
       adsurl = {https://ui.adsabs.harvard.edu/abs/2017A&A...605A..81D}
}

@ARTICLE{Furtak2024,
       author = {{Furtak}, Lukas J. and {Zitrin}, Adi and {Richard}, Johan and {Eckert}, Dominique and {Sayers}, Jack and {Ebeling}, Harald and {Fujimoto}, Seiji and {Laporte}, Nicolas and {Lagattuta}, David and {Limousin}, Marceau and {Mahler}, Guillaume and {Meena}, Ashish K. and {Andrade-Santos}, Felipe and {Frye}, Brenda L. and {Jauzac}, Mathilde and {Koekemoer}, Anton M. and {Kohno}, Kotaro and {Espada}, Daniel and {Lu}, Harry and {Massey}, Richard and {Niemiec}, Anna},
        title = "{A complex node of the cosmic web associated with the massive galaxy cluster MACS J0600.1-2008}",
      journal = {\mnras},
         year = 2024,
        month = sep,
       volume = {533},
       number = {2},
        pages = {2242-2261},
          doi = {10.1093/mnras/stae1943},
archivePrefix = {arXiv},
       eprint = {2404.03286},
 primaryClass = {astro-ph.GA},
       adsurl = {https://ui.adsabs.harvard.edu/abs/2024MNRAS.533.2242F}
}

@ARTICLE{Lagattuta2019,
       author = {{Lagattuta}, David J. and {Richard}, Johan and {Bauer}, Franz E. and {Cl{\'e}ment}, Benjamin and {Mahler}, Guillaume and {Soucail}, Genevi{\`e}ve and {Carton}, David and {Kneib}, Jean-Paul and {Laporte}, Nicolas and {Martinez}, Johany and {Patr{\'\i}cio}, Vera and {Payne}, Anna V. and {Pell{\'o}}, Roser and {Schmidt}, Kasper B. and {de la Vieuville}, Geoffroy},
        title = "{Probing 3D structure with a large MUSE mosaic: extending the mass model of Frontier Field Abell 370}",
      journal = {\mnras},
         year = 2019,
        month = may,
       volume = {485},
       number = {3},
        pages = {3738-3760},
          doi = {10.1093/mnras/stz620},
archivePrefix = {arXiv},
       eprint = {1904.02158},
 primaryClass = {astro-ph.GA},
       adsurl = {https://ui.adsabs.harvard.edu/abs/2019MNRAS.485.3738L}
}

@ARTICLE{Ho2012,
       author = {{Ho}, I. -Ting and {Ebeling}, Harald and {Richard}, Johan},
        title = "{An X-ray/optical study of the geometry and dynamics of MACS J0140.0-0555, a massive post-collision cluster merger}",
      journal = {\mnras},
         year = 2012,
        month = nov,
       volume = {426},
       number = {3},
        pages = {1992-2002},
          doi = {10.1111/j.1365-2966.2012.21806.x},
archivePrefix = {arXiv},
       eprint = {1207.6235},
 primaryClass = {astro-ph.CO},
       adsurl = {https://ui.adsabs.harvard.edu/abs/2012MNRAS.426.1992H}
}

@ARTICLE{Hsu2013,
       author = {{Hsu}, Li-Yen and {Ebeling}, Harald and {Richard}, Johan},
        title = "{The three-dimensional geometry and merger history of the massive galaxy cluster MACS J0358.8-2955}",
      journal = {\mnras},
         year = 2013,
        month = feb,
       volume = {429},
       number = {1},
        pages = {833-848},
          doi = {10.1093/mnras/sts379},
archivePrefix = {arXiv},
       eprint = {1209.2492},
 primaryClass = {astro-ph.CO},
       adsurl = {https://ui.adsabs.harvard.edu/abs/2013MNRAS.429..833H}
}

@ARTICLE{Christensen2012,
       author = {{Christensen}, Lise and {Richard}, Johan and {Hjorth}, Jens and {Milvang-Jensen}, Bo and {Laursen}, Peter and {Limousin}, Marceau and {Dessauges-Zavadsky}, Miroslava and {Grillo}, Claudio and {Ebeling}, Harald},
        title = "{The low-mass end of the fundamental relation for gravitationally lensed star-forming galaxies at 1 < z < 6}",
      journal = {\mnras},
         year = 2012,
        month = dec,
       volume = {427},
       number = {3},
        pages = {1953-1972},
          doi = {10.1111/j.1365-2966.2012.22006.x},
archivePrefix = {arXiv},
       eprint = {1209.0767},
 primaryClass = {astro-ph.CO},
       adsurl = {https://ui.adsabs.harvard.edu/abs/2012MNRAS.427.1953C}
}

@article{thai2023probing,
  title={Probing the faint-end luminosity function of Lyman-alpha emitters at 3< z< 7 behind 17 MUSE lensing clusters},
  author={Thai, Tran Thi and Tuan-Anh, Pham and Pello, Roser and Goovaerts, Ilias and Richard, Johan and Claeyssens, Ad{\'e}la{\"\i}de and Mahler, Guillaume and Lagattuta, D and De la Vieuville, Geoffroy and Salvador-Sol{\'e}, Eduard and others},
  journal={Astronomy \& Astrophysics},
  volume={678},
  pages={A139},
  year={2023},
  publisher={EDP Sciences}
}

@article{ebeling2017fully,
  title={Fully stripped? The dynamics of dark and luminous matter in the massive cluster collision MACSJ0553. 4- 3342},
  author={Ebeling, Harald and Qi, Jia and Richard, Johan},
  journal={Monthly Notices of the Royal Astronomical Society},
  volume={471},
  number={3},
  pages={3305--3322},
  year={2017},
  publisher={Oxford University Press}
}

@book{petters2012singularity,
  title={Singularity theory and gravitational lensing},
  author={Petters, Arlie O and Levine, Harold and Wambsganss, Joachim},
  volume={21},
  year={2012},
  publisher={Springer Science \& Business Media}
}

@article{oke1983secondary,
  title={Secondary standard stars for absolute spectrophotometry},
  author={Oke, JB and Gunn, JE},
  journal={Astrophysical Journal, Part 1, vol. 266, Mar. 15, 1983, p. 713-717.},
  volume={266},
  pages={713--717},
  year={1983}
}

@article{madau2014cosmic,
  title={Cosmic star-formation history},
  author={Madau, Piero and Dickinson, Mark},
  journal={Annual Review of Astronomy and Astrophysics},
  volume={52},
  number={1},
  pages={415--486},
  year={2014},
  publisher={Annual Reviews}
}

@article{jauzac2015hubble,
  title={Hubble Frontier Fields: a high-precision strong-lensing analysis of the massive galaxy cluster Abell 2744 using~ 180 multiple images},
  author={Jauzac, Mathilde and Richard, Johan and Jullo, Eric and Cl{\'e}ment, Benjamin and Limousin, Marceau and Kneib, J-P and Ebeling, H and Natarajan, P and Rodney, S and Atek, H and others},
  journal={Monthly Notices of the Royal Astronomical Society},
  volume={452},
  number={2},
  pages={1437--1446},
  year={2015},
  publisher={The Royal Astronomical Society}
}

@article{aazami2009universal,
  title={A universal magnification theorem for higher-order caustic singularities},
  author={Aazami, Amir B and Petters, Arlie O},
  journal={Journal of mathematical physics},
  volume={50},
  number={3},
  year={2009},
  publisher={AIP Publishing}
}

\begin{appendix}
\section{Cluster properties and exotic lensing diagnostics.}
Cluster properties and exotic lensing diagnostics. Summary of the lensing and structural parameters for all galaxy clusters analyzed in this study. The table lists the cluster redshift ($z_{\rm l}$), Einstein radius ($R_{\rm E}$), mass enclosed within $R_{\rm E}$ ($M$), ellipticity ($e$), compactness index ($\Delta\eta$), number of mass components within $R_{\rm E}$ ($N_{\rm pot}$), substructure mass fraction ($f_{\rm sub}$), projected exotic area ($\Omega$), and comoving exotic volume ($V$) for sources up to redshifts $z<4$ and $z<10$, together with the detection magnitude limit ($m_{\rm lim}$) and the expected number of hyperbolic umbilic source images ($N_{\rm gal}$). The exotic metrics $\Omega$ and $V$ are reported in the format $z<4$ ($z<10$). Clusters are ranked in descending order of exotic volume $V_{z<10}$. References: (1) \citet{postman2012cluster}, (2)\citet{Lotz2017}, (3) \citet{coe2019}, (4) \citet{Richard2021}, (5) \citet{richard2010}, (6) \citet{thai2023probing}, (7) \citet{ebeling2025}, (8) \citet{Lagattuta2023}, (9) \citet{repp2018}, (10) \citet{Dessauges2017}, (11) \citet{Furtak2024}, (12) \citet{Lagattuta2019}, (13) \citet{ebeling2017fully}, (14) \citet{Ho2012}, (15) \citet{Hsu2013}, and (16) \citet{Christensen2012}.
\longtab[1]{
\setlength\LTleft{\fill}
\setlength\LTright{\fill}
\begin{longtable}{@{\hspace{10pt}} l @{\hspace{10pt}} c @{\hspace{10pt}} c @{\hspace{10pt}} c @{\hspace{10pt}} c @{\hspace{10pt}} c @{\hspace{10pt}} c @{\hspace{10pt}} c @{\hspace{10pt}} c @{\hspace{10pt}} c @{\hspace{10pt}} c @{\hspace{10pt}} c @{\hspace{10pt}} c @{\hspace{10pt}}}
\label{tab:complete}\\
\hline
\hline
ClusterName & Ref & $z_l$ & $R_E$ & $M(r<R_E)$ & $e$ & $\Delta\eta$ & $N_{\rm pot}$ & $f_{\rm sub}$ & $\Omega_{z=4}\,(\Omega_{z=10})$ & $V_{z=4}\,(\,V_{z=10})$ & $m_{\rm lim}$ & $N_{\rm gal}$ \\
 &  &  & arcsec & $10^{14}\,M_{\odot}$ &  &  &  &  & arcsec$^2$ & Mpc$^3$ &  &  \\
\hline
\endfirsthead

\caption{Table~\ref{tab:complete}. continued.}\\
\hline
\hline
ClusterName & Ref & $z_l$ & $R_E$ & $M(r<R_E)$ & $e$ & $\Delta\eta$ & $N_{\rm pot}$ & $f_{\rm sub}$ & $\Omega_{z=4}\,(\Omega_{z=10})$ & $V_{z=4}\,(\,V_{z=10})$ & $m_{\rm lim}$ & $N_{\rm gal}$ \\
 &  &  & arcsec & $10^{14}\,M_{\odot}$ &  &  &  &  & arcsec$^2$ & Mpc$^3$ &  &  \\
\hline
\endhead

\hline
\multicolumn{13}{r}{\small\itshape (continued on next page)}\\
\endfoot

\hline
\endlastfoot
AS1063 &         1,2,4 & 0.347 & 30.22 & 1.17 & 0.50 & -0.02 & 44 & 0.079 & 15.82 (27.13) & 26.94 (104.94) & 29.0 & 0.881 \\
eMACS1527 &     7 & 0.697 & 20.73 & 0.799 & 0.64 & -0.61 & 11 & 0.012 & 20.77 (20.77) & 29.86 (88.28) & 27.5 & 0.349 \\
MACS1206 &     1,4,6 & 0.438 & 26.10 & 1.01 & 0.50 & -0.04 & 65 & 0.032 & 12.24 (13.91) & 22.18 (67.44) & 28.0 & 0.377 \\
MACS0417 &     3,6 & 0.443 & 21.51 & 0.689 & 0.58 & -0.45 & 3 & 0.016 & 9.66 (13.03) & 16.00 (59.41) & 27.5 & 0.172 \\
MACS0451 &     4,6 & 0.430 & 25.41 & 0.946 & 0.48 & -0.34 & 21 & 0.134 & 8.01 (18.58) & 11.01 (57.30) & 27.5 & 0.089 \\
RXJ0437+00 &     8 & 0.285 & 18.06 & 0.366 & 0.35 & 0.02 & 12 & 0.108 & 7.52 (9.56) & 14.98 (46.44) & 28.0 & 0.234 \\
eMACS0502 &     7 & 0.603 & 24.56 & 1.05 & 0.61 & -0.24 & 21 & 0.056 & 12.45 (12.45) & 24.18 (40.70) & 27.5 & 0.344 \\
MACS2214 &     4,6 & 0.502 & 18.61 & 0.553 & 0.59 & -0.42 & 10 & 0.054 & 8.43 (8.43) & 16.39 (40.51) & 28.0 & 0.295 \\
eMACS0256 &     7 & 0.862 & 10.91 & 0.239 & 0.42 & 0.20 & 6 & 0.027 & 4.89 (13.04) & 4.04 (40.10) & 27.5 & 0.033 \\
eMACS0252 &     7 & 0.702 & 15.47 & 0.446 & 0.59 & 0.06 & 10 & 0.112 & 5.68 (9.72) & 7.60 (39.02) & 27.5 & 0.063 \\
eMACS0934 &     7 & 0.563 & 16.43 & 0.457 & 0.59 & -0.22 & 9 & 0.041 & 6.02 (8.61) & 7.37 (35.85) & 27.5 & 0.074 \\
eMACS0840 &     7 & 0.638 & 13.92 & 0.347 & 0.30 & -0.18 & 4 & 0.004 & 6.10 (10.27) & 3.67 (34.54) & 27.5 & 0.034 \\
MACS0947 &     6,9 & 0.354 & 25.23 & 0.828 & 0.42 & -0.16 & 11 & 0.011 & 4.83 (12.86) & 5.23 (32.63) & 27.5 & 0.054 \\
eMACS1341 &     7 & 0.834 & 15.58 & 0.481 & 0.46 & 0.32 & 12 & 0.098 & 2.03 (14.19) & 1.76 (30.96) & 27.5 & 0.011 \\
MACS1731 &     6 & 0.389 & 25.36 & 0.887 & 0.76 & -0.22 & 10 & 0.260 & 4.82 (4.82) & 11.03 (27.33) & 27.5 & 0.154 \\
SMACS2332 &     4,6 & 0.398 & 16.50 & 0.381 & 0.62 & -0.17 & 7 & 0.095 & 4.55 (6.14) & 7.13 (27.23) & 27.5 & 0.083 \\
eMACS0121 &     7,13 & 0.601 & 11.99 & 0.251 & 0.51 & -0.48 & 5 & 0.011 & 6.34 (6.34) & 12.12 (26.64) & 27.5 & 0.14 \\
A1703 &         5 & 0.280 & 29.07 & 0.938 & 0.34 & 0.06 & 12 & 0.081 & 6.33 (6.33) & 13.09 (26.55) & 28.0 & 0.216 \\
MACS0257 &     4,6 & 0.322 & 25.45 & 0.791 & 0.39 & -0.07 & 36 & 0.177 & 5.31 (5.31) & 9.64 (26.26) & 27.5 & 0.157 \\
eMACS1248 &     7 & 0.573 & 16.69 & 0.476 & 0.50 & 0.14 & 6 & 0.026 & 3.45 (9.29) & 3.74 (23.79) & 27.5 & 0.042 \\
MACS0600 &     6,11 & 0.432 & 46.70 & 3.2 & 0.41 & -0.56 & 16 & 0.036 & 21.56 (21.56) & 17.74 (20.21) & 27.5 & 0.412 \\
eMACS1437 &     7 & 0.535 & 14.19 & 0.332 & 0.55 & -0.14 & 11 & 0.129 & 3.73 (4.04) & 5.63 (18.75) & 27.5 & 0.053 \\
MACS2050 &     6 & 0.333 & 17.53 & 0.384 & 0.17 & -0.03 & 8 & 0.011 & 3.50 (3.50) & 6.20 (17.99) & 27.5 & 0.062 \\
MACS0152 &     6 & 0.412 & 35.08 & 1.76 & 0.24 & -0.12 & 17 & 0.007 & 10.44 (10.44) & 14.36 (15.98) & 27.5 & 0.181 \\
A370 &         4,12 & 0.375 & 33.66 & 1.53 & 0.32 & 0.20 & 47 & 0.062 & 8.70 (8.70) & 12.18 (12.37) & 29.0 & 0.569 \\
MACS1311 &     3,6 & 0.494 & 18.28 & 0.529 & 0.33 & 0.22 & 4 & 0.028 & 1.56 (3.61) & 2.53 (12.17) & 27.5 & 0.027 \\
A2744 &         2,4 & 0.308 & 21.45 & 0.546 & 0.54 & 0.14 & 32 & 0.053 & 1.88 (2.18) & 3.53 (10.32) & 29.0 & 0.133 \\
MACS1532 &     1,6 & 0.345 & 24.82 & 0.788 & 0.21 & 0.06 & 36 & 0.045 & 2.10 (2.10) & 4.35 (9.67) & 27.5 & 0.056 \\
MACS1319 &     6 & 0.327 & 12.84 & 0.204 & 0.23 & 0.18 & 9 & 0.013 & 0.75 (4.43) & 0.85 (9.34) & 27.5 & 0.004 \\
eMACS2316 &     7 & 0.526 & 21.80 & 0.777 & 0.11 & -0.19 & 15 & 0.168 & 1.01 (2.49) & 1.23 (7.84) & 27.5 & 0.011 \\
MACS0329 &     1,4 & 0.450 & 24.11 & 0.874 & 0.22 & 0.23 & 26 & 0.107 & 0.79 (2.00) & 1.51 (7.46) & 27.5 & 0.021 \\
MACS0358 &     6 & 0.428 & 30.86 & 1.39 & 0.37 & -0.31 & 7 & 0.006 & 4.78 (4.78) & 6.29 (7.36) & 27.5 & 0.128 \\
MACS1931 &     6 & 0.345 & 22.36 & 0.64 & 0.76 & -0.62 & 10 & 0.066 & 2.41 (2.41) & 3.20 (6.88) & 27.5 & 0.049 \\
MACS1142 &     6 & 0.326 & 36.46 & 1.64 & 0.78 & -0.27 & 12 & 0.039 & 1.56 (1.56) & 2.69 (6.68) & 27.5 & 0.039 \\
eMACS0943 &     7 & 0.569 & 12.03 & 0.246 & 0.05 & 0.47 & 5 & 0.023 & 0.67 (3.35) & 0.43 (6.67) & 27.5 & 0.004 \\
MACS0032 &     6,10 & 0.377 & 23.97 & 0.778 & 0.49 & -0.10 & 0 & 0.053 & 1.17 (1.47) & 1.21 (6.21) & 27.5 & 0.006 \\
MACS0553 &     3,13 & 0.427 & 32.50 & 1.54 & 0.61 & -0.09 & 24 & 0.059 & 1.06 (2.35) & 1.96 (5.79) & 27.5 & 0.026 \\
MACS0520 &     6 & 0.336 & 26.98 & 0.915 & 0.31 & 0.48 & 17 & 0.065 & 1.67 (1.67) & 3.62 (5.79) & 27.5 & 0.052 \\
BULLET &         4 & 0.296 & 22.01 & 0.559 & 0.29 & 0.31 & 13 & 0.078 & 1.07 (1.95) & 1.84 (5.45) & 28.0 & 0.034 \\
MACS1720 &     6 & 0.391 & 21.80 & 0.658 & 0.30 & -0.24 & 12 & 0.112 & 2.32 (2.32) & 1.78 (5.44) & 27.5 & 0.029 \\
MACS1621 &     6 & 0.465 & 18.11 & 0.503 & 0.14 & -0.04 & 7 & 0.005 & 2.32 (2.32) & 3.83 (5.31) & 27.5 & 0.052 \\
MACS1452 &     6 & 0.324 & 28.02 & 0.964 & 0.04 & 0.31 & 14 & 0.088 & 0.38 (1.61) & 0.78 (5.14) & 27.5 & 0.008 \\
eMACS2229 &     7 & 0.621 & 10.62 & 0.2 & 0.77 & -0.15 & 7 & 0.001 & 0.34 (2.34) & 0.21 (5.10) & 27.5 & 0.001 \\
SMACS2031 &     4,6 & 0.331 & 21.38 & 0.569 & 0.30 & 0.23 & 17 & 0.031 & 1.09 (1.09) & 2.13 (4.92) & 27.5 & 0.034 \\
MACS0845 &     6 & 0.329 & 46.14 & 2.64 & 0.17 & 0.03 & 28 & 0.060 & 3.00 (3.00) & 3.12 (3.25) & 27.5 & 0.051 \\
MACS1752 &     6 & 0.364 & 11.19 & 0.166 & 0.37 & 0.41 & 7 & 0.093 & 0.58 (1.01) & 0.44 (3.13) & 27.5 & 0.005 \\
SMACS2131 &     6 & 0.442 & 19.16 & 0.547 & 0.23 & 0.12 & 29 & 0.294 & 0.42 (0.96) & 0.59 (2.94) & 27.5 & 0.006 \\
eMACS1852 &     7 & 0.604 & 14.53 & 0.369 & 0.45 & 0.14 & 12 & 0.029 & 1.26 (1.26) & 0.85 (2.19) & 27.5 & 0.004 \\
MACS0416 &     1,2,4 & 0.397 & 25.79 & 0.929 & 0.78 & 0.09 & 20 & 0.071 & 0.40 (0.40) & 0.74 (1.86) & 29.0 & 0.026 \\
eMACS0834 &     7 & 0.661 & 31.78 & 1.84 & 0.07 & -0.25 & 19 & 0.086 & 2.92 (2.92) & 1.67 (1.77) & 27.5 & 0.035 \\
A2667 &         4,5 & 0.233 & 17.69 & 0.304 & 0.36 & 0.07 & 10 & 0.040 & 0.23 (0.53) & 0.44 (1.73) & 28.0 & 0.006 \\
eMACS1209 &     7 & 0.555 & 22.08 & 0.819 & 0.05 & -0.51 & 52 & 0.034 & 0.31 (0.31) & 0.39 (1.21) & 27.5 & 0.006 \\
eMACS1057 &     7 & 0.603 & 20.30 & 0.72 & 0.20 & 0.14 & 15 & 0.124 & 0.04 (0.52) & 0.05 (1.06) & 27.5 & 0.0 \\
MACS1115\_00 &     1,6 & 0.352 & 60.50 & 4.74 & 0.29 & -0.76 & 19 & 0.004 & 18.58 (18.58) & 1.03 (1.03) & 27.5 & 0.0 \\
eMACS1756 &     7 & 0.574 & 13.44 & 0.309 & 0.81 & -0.65 & 0 & 0.045 & 0.10 (0.32) & 0.05 (0.85) & 27.5 & 0.001 \\
eMACS2327 &     7 & 0.706 & 35.14 & 2.31 & 0.53 & -0.91 & 20 & 0.004 & 6.17 (6.17) & 0.67 (0.80) & 27.5 & 0.023 \\
A2390 &         4,5 & 0.231 & 19.11 & 0.352 & 0.46 & 0.06 & 11 & 0.031 & 0.16 (0.18) & 0.31 (0.68) & 28.0 & 0.004 \\
eMACS0324 &     7 & 0.902 & 10.75 & 0.235 & 0.06 & 0.18 & 3 & 0.033 & 0.27 (0.27) & 0.09 (0.55) & 27.5 & 0.001 \\
MACS0712 &     6 & 0.328 & 5.85 & 0.0423 & 0.50 & 0.00 & 2 & 0.011 & 0.06 (0.13) & 0.12 (0.51) & 27.5 & 0.001 \\
eMACS0030 &     7 & 0.497 & 8.73 & 0.121 & 0.47 & 0.15 & 4 & 0.015 & 0.04 (0.20) & 0.05 (0.29) & 27.5 & 0.001 \\
eMACS1353 &     7 & 0.736 & 38.33 & 2.79 & 0.10 & 0.13 & 33 & 0.475 & 0.35 (0.35) & 0.12 (0.12) & 27.5 & 0.002 \\
MACS0140 &     6,14 & 0.451 & 21.59 & 0.702 & 0.63 & 0.05 & 17 & 0.026 & 0.11 (0.11) & 0.04 (0.04) & 27.5 & 0.001 \\
SMACS0304 &     6 & 0.460 & 9.12 & 0.127 & 0.19 & 0.19 & 4 & 0.014 & 0.04 (0.04) & 0.03 (0.04) & 27.5 & 0.0 \\
MACS1738 &     6 & 0.329 & 12.96 & 0.208 & 0.78 & -0.18 & 3 & 0.086 & 0.04 (0.04) & 0.02 (0.02) & 27.5 & 0.001 \\
MACS1133 &     6 & 0.389 & 11.31 & 0.176 & 0.40 & -0.05 & 8 & 0.074 & 0.06 (0.06) & 0.01 (0.01) & 27.5 & 0.0 \\
MACS0159 &     6 & 0.407 & 7.97 & 0.09 & 0.64 & -0.13 & 6 & 0.231 & 0.00 (0.00) & 0.00 (0.00) & 27.5 & 0.0 \\
SMACS0359 &     6 & 0.296 & 11.82 & 0.161 & 0.65 & -0.01 & 9 & 0.019 & 0.00 (0.00) & 0.00 (0.00) & 27.5 & 0.0 \\
MACS1226 &     6 & 0.437 & 6.00 & 0.0533 & 0.06 & -0.10 & 1 & 0.000 & 0.00 (0.00) & 0.00 (0.00) & 27.5 & 0.0 \\
eMACS1157 &     7 & 0.557 & 15.13 & 0.386 & 0.47 & 0.39 & 4 & 0.022 & 0.00 (0.00) & 0.00 (0.00) & 27.5 & 0.0 \\
eMACS1831 &     7 & 0.821 & 17.77 & 0.623 & 0.11 & -0.63 & 4 & 0.032 & 0.00 (0.00) & 0.00 (0.00) & 27.5 & 0.0 \\
MACS1354 &     6 & 0.397 & 7.33 & 0.0752 & 0.48 & 0.24 & 0 & 0.099 & 0.00 (0.00) & 0.00 (0.00) & 27.5 & 0.0 \\
MACS0940 &     6 & 0.335 & 9.94 & 0.124 & 0.30 & -0.09 & 11 & 0.277 & 0.00 (0.00) & 0.00 (0.00) & 27.5 & 0.0 \\
MACS2051 &     4,6 & 0.321 & 15.72 & 0.301 & 0.71 & -0.20 & 4 & 0.073 & 0.00 (0.00) & 0.00 (0.00) & 27.5 & 0.0 \\
MACS2140 &     6 & 0.313 & 17.51 & 0.368 & 0.21 & -0.03 & 6 & 0.012 & 0.00 (0.00) & 0.00 (0.00) & 27.5 & 0.0 \\
\end{longtable}
}
\end{appendix}
\end{document}